\documentclass[journal,10pt]{IEEEtran}
\usepackage{cite}
\usepackage{latexsym}
\usepackage{verbatim, amsbsy}
\usepackage{algorithm}
\usepackage{algpseudocode}

\usepackage{url}
\usepackage[caption=false]{subfig}
\usepackage{color,soul}
\usepackage{amsmath}                      
\usepackage{amssymb}
\usepackage{theorem}
\usepackage{amsfonts}
\usepackage{times}
\usepackage{graphicx}
\usepackage{enumitem}
\usepackage[normalem]{ulem}
\usepackage{bm}
\usepackage{array}
\usepackage{tikz}
\usetikzlibrary{arrows,positioning,shapes,decorations.pathmorphing,arrows.meta,math}
\usetikzlibrary{decorations.pathreplacing,patterns,intersections,backgrounds,calc}
\usepgflibrary{shapes.geometric}

\newcommand{\ben}{\begin{enumerate}}
\newcommand{\een}{\end{enumerate}}
\newcommand{\be}{\begin{equation}}
\newcommand{\ee}{\end{equation}}
\newcommand{\bea}{\begin{eqnarray}}
\newcommand{\eea}{\end{eqnarray}}
\newcommand{\bc}{\begin{cases}}
\newcommand{\ec}{\end{cases}}
\newcommand{\bi}{\begin{itemize}}
\newcommand{\Ei}{\end{itemize}}

\newcommand{\spa}{\,\,\,\!\!}

\newcommand{\mc}{\mathcal}

\newcommand{\Thr}{\Theta}

\newcommand{\Etx}{E_{\textrm{tx}}}
\newcommand{\xehd}{x_{\textrm{ehd}}}
\newcommand{\Pbo}{P_{\textrm{BO}}}
\newcommand{\Ntx}{N_{\textrm{tx}}}
\newcommand{\Nno}{N_{\textrm{no}}}
\newcommand{\Epla}{\varepsilon_{\textrm{plat}}}
\newcommand{\Epoi}{\varepsilon_{\textrm{pois}}}

\newcommand{\dpla}{d_0}

\algnewcommand{\algorithmicgoto}{\textbf{go to}}%
\algnewcommand{\Goto}[1]{\algorithmicgoto~\ref{#1}}

\algnewcommand{\algorithmicbreak}{\textbf{break}}%
\algnewcommand{\Break}[0]{\algorithmicbreak}

 \theoremstyle{break}

 {\theorembodyfont{\normalfont}     
 }

\def\sgn{\textrm{\ sgn}}

\def\de{\mathrm{d}}
\def\Ei{\textrm{Ei}}

\newcommand{\figww}{0.95\columnwidth} 

\begin{document}
\title{EH from V2X Communications: the Price of Uncertainty and the Impact of Platooning}

\author{Federico Librino,~\IEEEmembership{Member,~IEEE,} Paolo Santi~\IEEEmembership{Senior Member,~IEEE}
\thanks{Copyright (c) 20xx IEEE. Personal use of this material is permitted. However, permission to use this material for any other purposes must be obtained from the IEEE by sending a request to pubs-permissions@ieee.org.}
\thanks{F. Librino is with the Italian National Research Council, 56124 Pisa, Italy (e-mail: federico.librino@iit.cnr.it).}
\thanks{P. Santi is with the Italian National Research Council, 56124 Pisa, Italy and with the Massachussets Institute of Technology, 02139 Cambridge, MA, USA (e-mail: paolo.santi@iit.cnr.it).}
}

\date{August 2023}
\maketitle

\begin{abstract}
 In this paper, we explore how radio frequency energy from vehicular communications can be exploited by an energy harvesting device (EHD) placed alongside the road to deliver data packets through wireless connection to a remote Access Point. 
 Based on updated local topology knowledge, we propose a cycle-based strategy to balance harvest and transmit phases at the EHD, in order to maximize the average throughput.
 A theoretical derivation is carried out to determine the optimal strategy parameters setting, and used to investigate the effectiveness of the proposed approach over different scenarios, taking into account the road traffic intensity, the EHD battery capacity, the transmit power and the data rate. Results show that regular traffic patterns, as those created by vehicles platooning, can increase the obtained throughput by more than 30\% with respect to irregular ones with the same average intensity. 
 Black out probability is also derived for the former scenario. The resulting tradeoff between higher average throughput and lower black out probability shows that the proposed approach can be adopted for different applications by properly tuning the strategy parameters.
\end{abstract}


\section{Introduction}
In the next future, the Internet of Things (IoT) paradigm is expected to drastically increase the number of interconnected devices and networks, which will become a fundamental brick of the future smart cities.
A plethora of services and applications are envisioned to be made available to citizens everywhere and everytime, due to ubiquitous wireless connections. At the same time, Machine-to-Machine communications will play a pivotal role in order to gather and exchange relevant data in real time from streets, factories, vehicles, public offices and so on, so as to proactively take actions in order to keep all services available and efficient.

One of the key aspects of this scenario lies in its energy demand. While most of these devices will implement energy efficient protocols and transmission schemes when performing their communications, their sheer number will inevitably lead to a growing need for energy, which might offset the assumed carbon reduction goals for planned smart city transitions.

Recently, a promising solution to provide the required energy for communication devices has been identified in the exploitation of the energy scavenged from the ambient, where multiple sources can be properly leveraged. Several works have appeared in the literature, focusing on solar energy, vibrational energy, thermal energy and radio frequency (RF) energy.
Harvesting energy from RF transmissions is a particularly suitable option, since wireless communications will be part of most of the smart cities systems and applications. To this aim, the usage of dedicated wireless transmitters, as well as cellular networks base stations, has been explored.

In this work, we carry out a preliminary study on another potentially effective source of RF energy, namely the communications among vehicles. Indeed, vehicular communications are also foreseen to become pervasive with the development of smart mobility systems, as well as autonomous vehicles. In the future cities, the large majority of vehicles driving along the roads are expected to be exchanging data and information also through wireless channels, thus creating a valuable source of RF energy for devices located close to the main streets.

With the goal of achieving a tractable mathematical analysis, we investigate a simple yet significant scenario. We consider a single straight road segment where vehicles travel at constant speed. An EH device (EHD), willing to send its data to a sink node, is located close to the street, and is fed only by the RF energy of vehicular communications.
Our aim is to explore how the knowledge of the network topology, which we assume available, can be exploited to properly tune the EH device operational mode in order to maximize the communication throughput.
While limited, the obtained results can offer insights about the feasibility of the proposed approach, and a potential scheme to optimize the metrics of interest.

\subsection{Related Literature}
Energy harvesting has been widely investigated in the recent literature, since it provides an effective way to power mobile devices without the need for costly or inefficient battery replacing operations. The challenges in designing devices capable of scavenging energy have been addressed in early works~\cite{EH2,EH3}. Since then, several different sources of energy have been considered as feasible to power communication devices, ranging from solar energy~\cite{EH9,VEH18} to vibrational energy induced by the movement of vehicles in a highway~\cite{VEH13}. With the worldwide spread of wireless communications, Radio Frequency (RF) energy harvesting has also become a promising research direction, as reported in~\cite{EH7} and references therein. 
A common feature of all these energy sources lies in their inner stochastic nature. In order to faithfully reproduce the unpredictable fluctuations of energy availability, various stochastic models have been developed and utilized~\cite{Nico,Nico2,EH8,EH11,VEH18}. While a combination of multiple energy sources has been also envisioned~\cite{VEH16}, most of the existing work focus on harvesting from a single source.
In particular, RF energy harvesting has attracted more and more attention, since it makes the usage of dedicated energy sources viable.

Two main research directions can be distinguished in this field.
The first one focuses on simultaneous wireless information and power transfer (SWIPT) systems~\cite{VEH9,VEH11,VEH12,VEH20,VEH21,VEH22}.
Under this paradigm, power (or energy) transfer is intended as a specific application of RF energy harvesting, where the harvested energy is purposely emitted by a transmitter in order to convey part of its energy to the harvester.
Multiple-Input-Multiple-Output (MIMO) transmissions are often suitable for this type of approach. For instance, in~\cite{VEH9}, a node with a good channel towards the MIMO source harvests energy from the signal in order to forward it to another device with poorer channel conditions; the idea of EH powered relays is further developed in~\cite{VEH12}, where energy is accumulated or used depending on the channel conditions; in~\cite{VEH20}, the optimization of information and energy transmissions is found through a power splitting optimization algorithm in a scenario where Interference Alignment is also adopted.
The SWIPT approach can be implemented through either power splitting schemes or time splitting schemes, but it is feasible only when the source of energy and the source of data coincide.

The second research direction investigates wireless powered communication networks (WPCN)~\cite{VEH5,VEH10,VEH8,VEH7}. In this scenario, the source of data and the source of energy are distinct, with the latter that can be either dedicated or not.
In~\cite{VEH5}, RF energy is sent from a MIMO equipped Base Station over the downlink, so as to allow users to transmit data on the uplink, and the optimal BS transmit power is found as a function of the number of antennas and of users.
The idea of a hybrid Access Point acting as a power beacon is leveraged in~\cite{VEH10} as well. Here, throughput maximization and transmission time minimization problems are tackled in an IoT scenario, where the Short Packet Communication model is adopted.
In~\cite{VEH8} and~\cite{VEH9}, energy harvesting is used to power a cooperative jammer, which can protect the subsequent communication from an eavesdropper.
Note that in the abovementioned works, a dedicated source of energy is exploited. The main advantage of a power beacon is that the energy supply at the energy harvesting device can be guaranteed, at least within a certain level.

Nonetheless, harvesting energy from non dedicated, ambient sources is also feasible~\cite{VEH23,VEH24,VEH17}.
In~\cite{VEH23}, secondary users of a cognitive radio network harvest energy from ambient sources, including RF signals; authors derive the upper bound of the achievable throughput as a function of the energy availability, the primary traffic characteristics and the traffic detection threshold.
Harvesting energy from primary users transmissions in instead devised in~\cite{VEH24}. Stochastic geometry is used to determine the distribution of the transmitters, and the optimal density and transmission power of the secondary users is derived in order to maximize the secondary network throughput under constraints on the outage probability.
Authors in~\cite{VEH17} analyze the battery recharging time for a device that harvests energy from multiple RF sources, transmitting on different frequency channels, taking into account both large scale and small scale fading.
From the environmental viewpoint, the idea of reusing energy which is already available in the ambient is more appealing, thus explaining the research effort in this direction.

\subsection{Paper contributions}
Despite the vast amount of work in the literature, which have outlined several promising scenarios, to the best of our knowledge the usage of \emph{vehicle communications}~\cite{VEH25,VEH26,VEH27} as the main source of RF energy is yet an unexplored research direction. Such a scenario has some peculiar characteristics, the most important one being the partly predictable mobility patterns of the energy sources. This leads to the need for considerably different stochastic models able to capture the fluctuations of energy availability.
Motivated by this consideration, we propose a system where energy is scavenged from vehicle communications, which are foreseen to become pervasive in the next decades. The specific contributions of the article are:
\begin{itemize}
 \item the problem of finding the harvesting strategy that maximizes the throughput, based only on local topology information, is formalized mathematically, taking into account vehicular traffic statistics, battery capacity and power allocation;
 \item the optimization problem is reformulated by identifying a subclass of strategies that allow a cycle-based throughput maximization. The solution is derived analytically by modeling the battery status as a Markov Process;
 \item the \textit{price of uncertainty} on the achievable throughput and energy efficiency is shown by investigating different vehicular traffic statistics, highlighting the benefits of vehicles coordination techniques (e.g., platooning);
 \item a mathematical expression for the black-out probability is also carried out, and the tradeoff between this probability and the attainable throughput is also revealed.
\end{itemize}

The rest of the paper is organized as follows. Section~\ref{sec:sysmod} describes the system model, including the channel model and the energy handling model. Our proposed strategy is illustrated in Section~\ref{sec:strategy}, where the general maximization problem is formalized. The throughput mathematical derivation is carried out in Section~\ref{sec:throcomp}, and the results are presented in Section~\ref{sec:resu}. Finally, Section~\ref{sec:conclu} concludes the paper.
As to the notation, throughout the text bold variables, like $\mathbf{M}$, are used for matrices, while calligraphic symbols, like $\mc{S}$ or $\mc{C}$, indicate sets.

\section{System Model}
\label{sec:sysmod}
We consider a single lane of a straight road segment, where vehicles drive at a constant speed $v_0$. Their arrival follows a stochastic process, with the inter-vehicle distance $d_v$ distributed according to a general probability distribution function (pdf) $f_D(y)$. The vehicle density $\mu$ is defined as the reciprocal of $\mathbb{E}[d_v]$.
All the vehicles are equipped with a single antenna, and perform continuous wireless transmissions, with fixed power $P_v$, on dedicated channels, according to the 802.11p standard\footnote{Intermittent transmissions can be accounted for, by adding a transmission probability to the proposed model.}.

A transceiver capable of RF energy harvesting is placed at distance $w$ from the road. This EHD is powered by a rechargeable battery, with finite capacity $G$, and is willing to transmit data to an Access Point (AP) located at distance $r$ from it. We assume that the device is backlogged, and that it can recharge its battery by scavenging energy from the vehicular communications.
Furthermore, no information is known about the instantaneous battery status.
Although this agnostic system is certainly simplified, it can be applied to devices with minimal technology, and can still provide a performance lower bound for scenarios with more refined technology, where more complex control strategies can be envisioned.
Instead, local topology information is available, and includes, for each neighboring vehicle, its position and the currently employed wireless channel.
This information can be easily retrieved by the AP (or by the EHD itself) by listening to the broadcast channel utilized by vehicles to periodically exchange beacons. Beacons sent by a vehicle are very short packets, whose content includes the current position, speed and heading direction of the vehicle and its neighbors~\cite{myVT}, and are generally sent with relatively high frequency.
The AP may then inform the device about the expected vehicles movement through a dedicated downlink channel.

While the single lane scenario under investigation is clearly a simple one, we consider this work as a first brick towards more complex scenarios: multiple lanes and/or streets, and variable vehicle speeds will require to take into account more complex topologies, with time-varying inter-vehicle distances, while the presence of intersections might induce different but partly predictable traffic patterns. At the same time, the deployment of multiple EHDs will allow the development of network protocols, able to exploit the inhomogeneous spatial RF energy distribution to smartly identify multi-hop routes able to deliver packets where they are needed.
Data relaying by vehicles may be also envisioned as an alternative approach. Indeed, the vehicle can leverage a battery with much higher capacity, and could therefore relay the data from the EHD to the AP. There may be, however, some drawbacks in this solution. Firstly, the vehicles are assumed to be moving, while the EHD has a fixed (and probably smartly planned) position. This makes it possible to employ features able to improve the EHD-AP channel (e.g., directional antennas, or precise beamforming through MIMO techniques at the AP, which would also result in better resilience against interference from multiple EHDs).
Secondly, the two-hop transmission from the EHD to a vehicle and then to the AP should be performed on one of the channels used by the vehicles for V2X communications. This may be impractical in an urban scenario with high vehicular density, where these channels are likely to already sustain a high load. Moreover, a coordination protocol should be implemented in order to avoid multiple vehicles to forward the data from the same EHD, thus further increasing the channels load due to the necessary overhead.
We leave these interesting extensions as promising future research directions.

We consider a slotted time model, where the slot duration is $T$.
The EHD is equipped with a single antenna, that is used for both transmission and energy harvesting. These two operations are mutually incompatible, meaning that, at a given time slot, the EHD can be either in \emph{Transmit} or in \emph{Harvest} mode. The energy scavenged in the Harvest mode is used to replenish the battery, which later provides this energy for data transmission.
All the relevant parameters are listed in Table~\ref{tab:param}.
\begin{table}
 \centering
 \begin{tabular}{cl}
  \hline\hline
  Notation & Parameter\\
  \hline $d_v$ & Inter-vehicle distance [m]\\
  $\mu$ & Intensity of arrivals Poisson process [vehicle/m]\\
  $r$ & Distance between EHD and AP [m] \\
  $w$ & Distance between EHD and lane center [m]\\
  T & Duration of a time slot [s]\\
  $v_0$ & Vehicles speed [m/s]\\
  $P_t$ & EHD transmit power [W]\\
  $P_v$ & Vehicle transmit power [W] \\
  $N_0$ & Noise power [W]\\
  $\alpha$ & Path loss coefficient \\
  $\kappa$ & Rice factor \\
  $\eta$ & Harvesting efficiency\\
  $\Etx$ & Energy required for data packet tx [J]\\
  $B$ & Channel bandwidth between EHD and AP [Hz]\\
  $S$ & Data packet size [bit]\\
  $\phi_s$ & Decoding probability at AP\\
  $G$ & Battery capacity [J]\\
  $N_s$ & Energy quanta in a full battery\\
  $\ell$ & Harvesting distance [m]\\
  \hline\hline\\
 \end{tabular}
 \caption{List of parameters.}
 \label{tab:param}
 \vspace{-0.8cm}
\end{table}

\subsection{EHD transmissions model}
When the EHD device is in Transmit mode, it sends one data packet per slot to the AP, with transmit power $P_t$. The energy required to send a packet is equal to $E_{\textrm{tx}} = P_tT$, which is drained from the battery. If the battery level is already below $\Etx$, the transmission is unsuccessful, and the battery is completely discharged; if the battery is already empty, the EHD remains silent.
We assume that data packets have size $S$ bit. The channel model from the EHD to the AP includes both path loss, with exponent $\alpha$, and Rayleigh quasi-static block fading. Hence, a data packet sent at time slot $n$ is received at the AP with a Signal-to-Noise ratio given by
\begin{equation}
 \textrm{SNR}(n) = \frac{P_tr^{-\alpha}}{N_0}|h(n)|^2,
 \label{SNR}
\end{equation}
where $N_0$ is the noise power, and $h(n)$ is modeled as a complex Gaussian random variable with zero mean and unit variance. Therefore, $|h(n)|^2$ is an exponential random variable with unitary mean. The fading coefficients are considered i.i.d. across time. According to the Shannon capacity model, the packet is correctly received if
\begin{equation}
 B\log_2(1+\textrm{SNR}(n)) \geq \frac{S}{T},
\end{equation}
where $B$ is the channel bandwidth, thus yielding the decoding probability
\begin{equation}
 \phi_s = \exp\left(-\frac{N_0r^{\alpha}}{P_t}\left(2^{\frac{S}{BT}}-1\right)\right).
 \label{decprob}
\end{equation}

\subsection{EHD harvesting model}
\label{sec:harvmodel}
When the EHD is in Harvest mode, it uses its antenna to scavenge energy from the neighboring vehicles. We consider a linear model for the harvesting, that is, a fraction $\eta$ of the incoming energy is actually stored in the battery.
Non linear harvesting model has been investigated as well\cite{VEH2}, since they are closer to the behavior of real devices. Nevertheless, as shown in \cite{VEH1}, the difference between the two models becomes relevant when the input power is high, e.g., in the presence of power beacons specifically deployed to supply energy to surrounding EHDs.
In our scenario, instead, energy is obtained from vehicular communications, whose adopted power is much lower: even when the vehicle is closest to the EHD, the input power $P_v/w^{\alpha}$ is lower than 1 mW, thus the linear model can be chosen as a tight approximation of the more realistic exponential one. 

Following the abovementioned channel model, the incoming power from a vehicle at distance $d_t$ at time instant $t$ is
\begin{equation}
 P_r(t) = \frac{P_v}{d_t^{\alpha}}|h_v(t)|^2.
 \label{recpow}
\end{equation}
For the fading coefficient $h_v(t)$, we consider Rician fading, which more faithfully reproduces the LOS communications occurring when the EHD is closer to the street, and is often more suitable for short range transmissions~\cite{rician}. In this case, $h_v(t)$ is modeled as a Rice random variable with scale parameter equal to 1 and Rice factor (that is, the ratio between the power of the LOS component and the scattered ones) equal to $\kappa$.
The amount of energy harvested from this vehicle (and stored in the battery) on a time interval $\Delta$ centered at $t$ is therefore $\Delta\eta P_r(t)$.
Notice that this holds only if $\Delta$ is small, since both the channel fading and the vehicle position change across time.
This makes the modeling of harvested energy less straightforward. On one side, averaging out the fading would allow a continuous time model, able to better represent the continuous position variation.
Such a model, while offering some general insights on the overall system evolution, would miss the impact of the inherent stochastic nature of the wireless channel.
On the other side, it is possible to rely on a discrete time model, such that in any time step the vehicle position and the channel fading are considered fixed, with fading coefficients independent across time.
The choice of the time step size must be done carefully. Too long time steps would not capture the vehicles movement adequately, especially when vehicle speed is high; conversely, too short time steps would make the hypothesis of independent fading coefficients unrealistic, thus requiring to take into account involved stochastic correlations.
In this work, for the sake of simplicity, we set the quantization step equal to the time slot, which was purposely chosen so as to guarantee i.i.d. fading coefficients.
The scavenged energy on a given slot $n$ is hence $\eta P_r^{(n)}T$, where $P_r^{(n)}$ is obtained from (\ref{recpow}) by using the fading coefficient $h_v(n)$ and setting $d_t$ equal to the distance between the EHD and the midpoint of the road segment traversed by the vehicle during that time slot.
This energy is stored in the battery, up to its capacity $G$: if additional energy is obtained, it cannot be stored, and is consequently lost.

When multiple vehicles are present, they will transmit on different channels of a given spectral band (e.g., the 5.9 GHz band for 802.11p) in order to avoid interference, especially if they are close to each other. The EHD is assumed to be designed to operate on the entire fraction of spectrum dedicated to V2X communications, and to be therefore able to perform energy harvesting irrespective of the specific channel selected by each vehicle.

\subsection{EHD battery model}
The energy harvested by the EHD is stored in a finite capacity battery, which then provides the energy required for data transmission.
Call $\Phi_n$ the battery level at the beginning of time slot $n$. If the EHD is in Transmit mode, then the battery status updates as
\begin{equation}
 \Phi_{n+1} = \max(\Phi_n-\Etx, 0).
\end{equation}
Conversely, if it is in Harvest mode, the update is
\begin{equation}
 \Phi_{n+1} = \min(\Phi_n + \eta P_r^{(n)}T, G),
\end{equation}
where $P_r^{(n)}$ is the incoming power in the current time slot, assuming fixed vehicles positions and fading coefficients, as explained above.
While the energy expense in the Transmit mode is inherently quantized, due to the fixed transmit power level $P_t$, the same does not hold for the energy harvested in the Harvest mode, since $P_r^{(n)}\in\mathbb{R}^+$.
As stated above, however, no information about the current battery status $\Phi(n)$ is available at the EHD.

\subsection{Throughput definition and problem formulation}
The throughput is defined as the number of packets correctly delivered over a given time interval. Call $s_n\in\{0,1\}$ the mode selected by the EHD at time slot $n$, which can be either Transmit mode ($s_n=1$) or Harvest mode ($s_n=0$). In time slot $n$, a packet is delivered if the following conditions hold:
\begin{itemize}
 \item $s_n = 1$, that is, the EHD selects Transmit mode;
 \item $\Phi_n\geq\Etx$, that is, there is enough energy in the battery;
 \item the packet is correctly decoded at the destination.
\end{itemize}
The third condition does not depend on the EHD operational mode, but only on the channel conditions, and can be modeled through the delivery probability $\phi_s$ in (\ref{decprob}). The first condition depends on the EHD mode of the present time slot, and the second one depends (also) on the operational mode selected in the previous time slots.
Hence, the throughput is defined as
\begin{equation}
 \Thr = \lim_{N\rightarrow+\infty}\frac{\phi_s}{NT}\sum_{n=1}^N\chi(\Phi_n\geq\Etx)s_n,
 \label{throgen}
\end{equation}
where $\chi(\cdot)$ is the indicator function. From (\ref{throgen}), it follows that the throughput depends on the operational mode selection of the EHD at each time slot since, even when Transmit mode is chosen, a successful transmission can be performed only if there is enough energy stored in the battery.

The choice of the operational mode in a slot must be based on the local topology, since this is the only available information. Call $\mathcal{C}$ the set of all the possible local topology configurations, each one characterized by the number and locations of the vehicles within a given distance from the EHD. A \emph{strategy} $\sigma:\mathcal{C}\rightarrow\{0,1\}$ maps any possible local topology configuration into a binary value that determines the operational mode. Hence, the set $\mathcal{H}$ of all possible strategies has cardinality $2^{|\mathcal{C}|}$. Under strategy $\sigma$, the value of $s_n$ at slot $n$ is set to $\sigma(C_n)$, being $C_n\in\mathcal{C}$ the topology configuration at that time slot.

The related throughput maximization problem can hence be written as
\begin{equation}
 \begin{aligned}
  \underset{\sigma\in\mathcal{H}}{\max} \quad & \mathbb{E}\left[\lim_{N\rightarrow+\infty}\frac{\phi_s}{NT}\sum_{n=1}^N\chi(\Phi_n\geq\Etx)\sigma(C_n)\right] \\
  \text{s. t.} \quad & C_n\in\mathcal{C}, \\
  & \Phi_n \in[0,G], \\
  & \Phi_n = \Phi_{n-1} + \min(\eta P_r^{(n)}T, G-\Phi_{n-1})(1-\sigma(C_n))+\\
  & \hspace{.8cm} - \min(\Etx,\Phi_{n-1})\sigma(C_n).
 \end{aligned}
 \label{optprob}
\end{equation}
Notice that problem (\ref{optprob}) is hard to tackle: $P_r^{(n)}$ is known only stochastically, based on the vehicles position at each slot, and the maximization is hence done over the expected value of $\Thr$. Moreover, the size of $\mathcal{C}$ is potentially very high, making the number of possible strategies overwhelming.

\section{Cycle-based mode selection strategy}
\label{sec:strategy}
In this section, we propose a heuristic approach to solve (\ref{optprob}), based on two observations. The former is that, due to the law of energy propagation, even when there are multiple vehicles in the proximity of the EHD, the energy harvested by the closest one is by far the largest fraction.
Hence, we can characterize a topology configuration by simply using the (quantized) distance of the closest vehicle from the EHD, which lets us consistently reduce the cardinality of set $\mathcal{C}$ (and hence of set $\mathcal{H}$).

The second observation is that, while the selection of the Transmit mode gives a reward that is stochastically independent from the time slot (since both the EHD and the AP remain at the same location, thus leaving the decoding probability unaltered across time), the same is not true for the Harvest mode. Indeed, choosing the Harvest mode is particularly advantageous only when there is a vehicle close to the EHD location, thus reducing the impact of path loss and increasing the expected value of harvested energy. The key idea is hence to adopt Harvest mode as long as a vehicle is close enough, and Transmit mode otherwise.
By doing this, we are actually restricting the search of the optimal strategy to the subset $\hat{\mathcal{H}}\subset\mathcal{H}$ containing only the strategies that prescribe Harvest Mode when the closest vehicle is nearer than a threshold distance, which we call \emph{harvest distance}, and Transmit Mode otherwise. 

This approach effectively divides time into \emph{cycles}, each composed by a \emph{Harvest Phase} (HP) followed by a \emph{Transmit Phase} (TP). During the time slots of the former phase, Harvest mode is always selected, while Transmit mode is chosen in all the time slots of the latter phase.
A cycle begins when a vehicle gets closer to the EHD than the harvest distance, thus triggering a new HP, which lasts until the vehicle moves farther than the same threshold. At this point, the TP starts, and continues until the next vehicle gets closer than the harvest distance, which corresponds to the start of the next cycle\footnote{The TP might be absent when two vehicles are very close to each other, since the latter might get closer than the harvest distance when the former is still within the same distance. Conversely, over a long TP, the EHD might run out of energy, thus remaining silent until the following HP.}.

Each strategy in $\hat{\mathcal{H}}$ is characterized by a different harvest distance value. Therefore, finding the best solution is equivalent to finding the optimal harvest distance.
Making it large grants more time to recharge the battery; however, any time slot in the HP corresponds to a missed transmission opportunity, and inflating it too much has a detrimental effect. Conversely, reducing the threshold leaves more time for transmissions, but at the cost of a lower average amount of harvested energy, thus increasing the risk of an energy outage.
Notice that the harvest distance might be limited by the EHD sensitivity, that is, the minimum amout of incoming power necessary for actual harvesting. While we do not model sensitivity explicitly, it can be easily included by adding a constraint to the harvest distance, which depends on the specific device implementation. For instance, with a sensitivity of -18.5 dBm, which is already achieved in some commercial products, we get that the harvest distance should be lower than 18 m, which will be proved to be much higher than the optimal value in the considered scenario.

\subsection{Simplified Problem Formulation}
Consider a Cartesian plane such that the considered road lane is placed along the $x$ axis, with the vehicles moving eastward. The EHD is placed at coordinates $(\xehd,-w)$. For mathematical tractability, the harvest distance $\ell$, triggering the beginning of a new cycle, is not measured between the vehicle and the EHD, but between the vehicle and the projection of the EHD location onto the road segment, that is, point $(\xehd,0)$. The corresponding distance between the vehicle and the EHD can be easily derived.

\begin{figure}
\centering
 \begin{tikzpicture}[>=stealth,scale=.58]
  \def\lm{0.8}; 
  \def\shr{3}; 
  \def\w{3}; 
  \def\iv{5}; 
  \def\ehdx{7}; 
  \def\apx{2}; 
  
  \fill[fill=gray!50] (-2,-0.8) rectangle +(14,1.8);
  \draw[dashed,white,line width=.1cm, dash pattern=on 0.8cm off 0.8cm] (-2,0.1) -- (12,0.1);
  \draw[thick] (-2,-0.8) -- (12, -0.8);
  \draw[thick] (-2,1) -- (12, 1);
  
  \foreach \vc in {0,\iv,...,12}
  {
    \node[inner sep=0pt] (car1) at (\vc,-0.35) {\includegraphics[height=0.4cm]{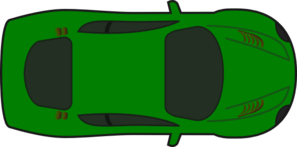}};
  }
  \node at (-0.6,0.3) {$V_{q+1}$};
  \node at (\iv+0.6,0.3) {$V_q$};
  
  \node[inner sep=0pt] (wsens) at (\ehdx,-1.5) {\includegraphics[height=0.7cm]{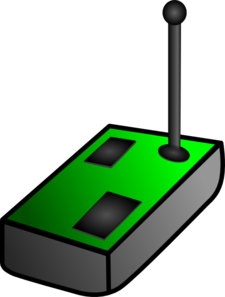}};
  \node[inner sep=0pt] (AP) at (\apx,-2.5) {\includegraphics[height=1cm]{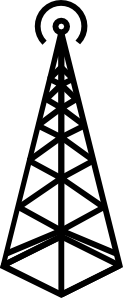}};
  
  \draw[thick,red,<->] (0, 0.4) -- (\iv, 0.4) node[midway,yshift=0.2cm] {$d_q$};
  \draw[thick,red,<->] (\ehdx+.5, -1.5) -- (\ehdx+.5, -0.35) node[midway,xshift=0.2cm,yshift=-0.1cm] (6.3,-0.75) {$w$};
  \draw[thick,red,<->] (\ehdx, -2) -- (\apx, -3.2) node[midway,yshift=0.2cm] {$r$};

  \draw[dashed] (\ehdx, -0.35) -- (\ehdx,1.7);
  \node at (\ehdx,-0.35) [circle,fill,inner sep=1.2pt]{};
  \node at (\ehdx,2) {$\xehd$};
  
  \node (xpl) at (\ehdx+\lm+2,1.8) {$\xehd\!+\!\ell$};
  \node at (\ehdx+\lm,-0.35) [circle,fill,inner sep=1.2pt]{};
  \draw[dashed] (\ehdx+\lm, -0.35) -- (\ehdx+\lm,1.4) -- (xpl.west);
  
  \node (xpl) at (\ehdx-\lm-2,1.8) {$\xehd\!-\!\ell$};
  \node at (\ehdx-\lm,-0.35) [circle,fill,inner sep=1.2pt]{};
  \draw[dashed] (\ehdx-\lm, -0.35) -- (\ehdx-\lm,1.4) -- (xpl.east);
  
  \draw[thick,->] (-2,-0.35) -- (12.8,-0.35) node[below] {$x$};
  \draw[thick,->] (-1.8,-2) -- (-1.8,2) node[right] {$y$};
  
  \begin{scope}[shift={(0,4)}]
  \clip(-2,-0.8) rectangle +(15.2,3.2);
  \foreach \vc in {0,\iv,...,12}
  {
    \fill [yellow] (\vc+\lm,0) -- (\vc-\lm,0) -- plot[domain=\vc-\lm:\vc+\lm]  (\x, {50/((\shr*(\x-\vc))^2 + \w^2)^1.5});
    \draw[thick,color=red,domain=\vc-\iv/2:\vc+\iv/2] plot (\x, {50/((\shr*(\x-\vc))^2 + \w^2)^1.5});
  }
  \draw[thick,->] (-2,0) -- (12.8,0) node[below] {$x$};
  \draw[thick,->] (-1.8,-0.2) -- (-1.8,2) node[right] {$P$};
  
  \draw[blue,<->] (\iv-\lm, -0.2) -- (\iv+\lm, -0.2) node[midway,yshift=-0.2cm] {$2\ell$};
  \draw[red,<->] (\lm, -0.2) -- (\iv-\lm, -0.2) node[midway,yshift=-0.2cm] {$d_q-2\ell$};
  \end{scope}
  
 \end{tikzpicture}
 
 \caption{Graphic representation of the considered scenario. Above, a qualitative graph of the average received power as a function of the vehicles positions is also reported, which shifts as the cars move. Since the EHD harvests energy when the x-coordinate of the closest vehicle is within the interval $[\xehd-\ell,\xehd+\ell]$, the yellow shaded area is proportional to the harvested energy.}
 \label{fig:gramodel}
\end{figure}
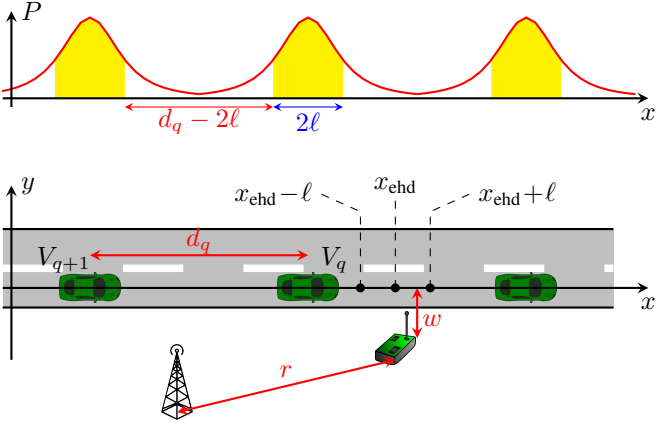

The $q$--th cycle starts when vehicle $V_q$ arrives at position $(\xehd-\ell,0)$, and lasts until the following vehicle $V_{q+1}$ gets to the same location. The HP starts at the beginning of the cycle, and ends when $V_q$ arrives at position $(\xehd+\ell,0)$.
Figure~\ref{fig:gramodel} gives a picture of the considered scenario. In the upper part, a qualitative sketch of the average amount of incoming power, as a function of the vehicles positions, is also depicted.

Since the speed is assumed constant, in the $q$--th cycle the HP lasts for $N_H = 2\ell/(v_0 T)$ time slots, while the TP lasts for $N_T(q) = (d_q-2\ell)/(v_0T)$ time slots, being $d_q$ the distance between $V_q$ and $V_{q+1}$.
It must be observed, however, that not all the $N_T(q)$ transmissions might actually take place, since energy outage events might occur. Call $\Phi(q)$ the energy stored in the battery at the beginning of cycle $q$, and $E_h(q)$ the energy collected during the HP of cycle $q$. The number $W_q$ of transmissions performed in cycle $q$ becomes
\begin{equation}
 W_q = \min\left(N_T(q),\frac{\min\left(\Phi(q)+E_h(q),G\right)}{\Etx}\right),
 \label{defW}
\end{equation}
where the maximum theoretical number of transmissions $N_T(q)$ is limited by the amount of transmissions that can be performed with the available energy. This is in turn given by the sum of the energy $\Phi(q)$ already present in the battery and the one harvested during the HP.

Leveraging the i.i.d. property of the inter-vehicle distances, we can model the system as a Renewal-Reward process, where the reward variable $W_q$ is the number of transmissions performed in the $q$--th cycle, while the holding time $Z_q$ is the duration of the cycle, equal to $d_q/v_0$.

Notice that while the $Z_q$'s are i.i.d. by assumption, the $W_q$'s are identically distributed, but in general not independent.
The source of correlation among $W_q$ and $W_{q+1}$ lies in the energy stored in the battery at the end of the $q$--th cycle, which induces a negative correlation. In fact, a high number of transmissions in cycle $q$ means that lower energy is left at the beginning of cycle $q+1$, resulting in a stochastically lower value for $W_{q+1}$.
This correlation mainly depends on the battery capacity; however, it is quite loose, as confirmed by the accuracy of the obtained results, and can be neglected.

According to the elementary renewal theorem for Renewal-Reward processes, the average throughput can be written as
\begin{equation}
 \Thr = \frac{\mathbb{E}\left[W_q\right]}{\mathbb{E}[Z_q]},
\end{equation}
and the throughput maximization problem can be reformulated simply as
\begin{equation}
\underset{\ell}{\max} \quad \frac{\phi_s v_0}{\mathbb{E}[d_v]} \mathbb{E}_{\Phi,E_h,d_v}\left[\min\left(\frac{d_v-2\ell}{v_0T},\frac{\min(\Phi+E_h,G)}{\Etx}\right)\right],
 \label{optsim}
\end{equation}
where the number of transmissions must be averaged over the inter vehicle distance, the amount of energy in the battery and the amount of harvested energy during the HP.
We compute the throughput as
\begin{eqnarray}
 \Thr\!\!\!\! & = \!\!\!\! & \frac{\phi_sv_0}{\mathbb{E}[d_v]}\int_0^G\int_0^{+\infty}\!\!\int_0^{+\infty}\!\!\!\min\left(\frac{y-2\ell}{v_0T},\frac{\min(z+x,G)}{\Etx}\right)\!\times \nonumber \\
 & & \times\spa f_E(x)f_D(y)f_B(z)\de x\de y\de z.
 \label{optint}
\end{eqnarray}
The distribution of the inter vehicle distance $d_v$ is given by the generic distribution $f_D(\cdot)$, while $f_E(\cdot)$ is the distribution of the per cycle harvested energy $E_h$, and $f_B(\cdot)$ is the distribution of the battery level at the beginning of a cycle. The derivations of $f_E(\cdot)$ and $f_B(\cdot)$ are detailed below.

\subsection{Distribution of the per cycle harvested energy $E_h$}
The HP takes place in the time interval during which the vehicle $V_q$ that provides energy traverses a segment of length $2\ell$. According to the discrete model outlined above, we divide this road segment into $L$ segments of length $v_0T$, each one covered by the vehicle in a single time slot.

Its movement is quantized by assuming that, during each time slot, the vehicle is fixed at the midpoint of the crossed segment.
The energy $E_h$ harvested then reads as
\begin{equation}
 E_h = \sum_{i=0}^{L-1} \frac{\eta P_vT}{\left[w^2 + \left(\left(i+\frac{1}{2}\right)v_0T-\ell\right)^2\right]^{\alpha/2}}X_i = \sum_{i=0}^{L-1}\frac{X_i}{\lambda_i},
 \label{segdiv}
\end{equation}
with $L = \lceil2\ell/(v_0T)\rceil$, and
\begin{equation}
 \lambda_i =  \frac{1}{\eta P_vT}\left[w^2 + \left(\left(i+\frac{1}{2}\right)v_0T-\ell\right)^2\right]^{\alpha/2},
\end{equation}
where $X_i$ represents the square of the rician fading coefficient experienced in the $i$--th road segment, and is therefore equal to $Y_i/(2(\kappa+1))$, being $\kappa$ the Rice factor and $Y_i$ a noncentral chi-square random variable with 2 degrees of freedom and non centrality parameter $2\kappa$ (thus correctly resulting in $\mathbb{E}[X_i] = 1$). Hence, we can rewrite the overall harvested energy as
\begin{equation}
 E_h = \frac{1}{2(\kappa+1)}\sum_{i=0}^{L-1}\frac{Y_i}{\lambda_i},
 \label{ricecomb}
\end{equation}
which is basically a linear combination of noncentral chi-square random variables.
The resulting distribution is quite involved, and has no closed form expression. We then approximate it using the saddle point approximation as follows. The moment generating function of $E_h$ is
\begin{equation}
 M_E(t) = \prod_{i=0}^{L-1}\frac{\hat\lambda_i}{\hat\lambda_i-2t}\exp\left(\frac{2\kappa t}{\hat\lambda_i-2t}\right),
\end{equation}
where $\hat\lambda_i=2(\kappa+1)\lambda_i$.
The cumulant generating function $K_E(t)=\ln(M_E(t))$ is therefore
\begin{equation}
 K_E(t) = \sum_{i=0}^{L-1}\left(\ln\left(\frac{\hat\lambda_i}{\hat\lambda_i-2t}\right)+\frac{2\kappa t}{\hat\lambda_i-2t}\right).
\end{equation}
Its first and second derivatives are, respectively,
\begin{eqnarray}
 K'_E(t) & = & 2\sum_{i=0}^{L-1}\frac{1}{\hat\lambda_i-2t}\left(1+\frac{\kappa\hat\lambda_i}{\hat\lambda_i-2t}\right), \\
 K''_E(t) & = & 4\sum_{i=0}^{L-1}\frac{1}{(\hat\lambda_i-2t)^2}\left(1-4\kappa\frac{\hat\lambda_i-3t}{\hat\lambda_i-2t}\right).
\end{eqnarray}
The CDF $F_E(x)$ can thus be obtained as~\cite{lugarice,approRV}
\begin{equation}
 F_E(x) = \Phi(u) + \frac{e^{-u^2/2}}{\sqrt{2\pi}}\left(\frac{1}{u}-\frac{1}{v}\right),
 \label{ecdf2}
\end{equation}
where $\Phi(\cdot)$ is the CDF of the standard normal distribution, $\hat z$ is the solution of the saddle point equation $K'_E(\hat z) = x$, while $u = \sgn(\hat z)\sqrt{2(\hat zx-K_E(\hat z))}$ and $v=\hat z\sqrt{K_E''(\hat z)}$.

\subsection{Battery status statistics}
\label{sec:batstat}
Differently from the harvested energy $E_h$, whose distribution is independent across cycles, the battery status evolves across time, and a different approach is needed in order to cope with the resulting time correlation.
The evolution of the battery status on a cycle basis is given by the following expression:
\begin{equation}
 \Phi(q+1) = \max\left(\min\left(\Phi(q)+E_h(q),G\right) - \Etx N_T(q), 0\right),
\end{equation}
where $N_T(q)$ is the number of slots in the TP during cycle $q$, $E_h(q)$ is the harvested energy in the HP, and $\Phi(q)$ is the battery charge at the beginning of cycle $q$.
Since the battery status depends only on its value in the previous cycle, we can model its evolution as a discrete time Markov Process, whose status $s_q$ at instant $q$ is given by the battery level at the beginning of cycle $q$.

In principle, the battery level can assume any real value in the interval $[0,G]$. For mathematical tractability, we introduce an approximation by quantizing it with quantization step $\Etx$. We hence assume that the harvested energy $E_h$ in the HP is rounded down to the closest multiple of $\Etx$. Since $\Etx$ is usually much smaller than $G$, we expect this approximation to have negligible effects on the overall system performance.

Let us define $N_s = G/\Etx$, which means that a fully charged battery can provide energy for at most $N_s$ data packet transmissions (independently from the packet size $S$). According to the quantization, the Markov process is turned into a discrete state process, with the states given by the set $\mathcal{S} = \{0,\Etx,2\Etx,\ldots,N_s\Etx\}$.
Coherently, the battery status distribution $f_B(\cdot)$ in (\ref{optint}) is replaced by a discrete probability mass function $p_B(\cdot)$, defined over the domain $\mathcal{S}$, and equal to the status steady state distribution $\mathbf{\pi}$ of the Markov process.

In order to find $\mathbf{\pi}$, we first derive the Markov process transition matrix $\mathbf{M}$. The transition from state $s_q\in\mathcal{S}$ to state $s_{q+1}\in\mathcal{S}$ depends on both the energy harvested during the HP and the energy depleted during the TP.
As to the former, the probability mass function $p_E(\cdot)$ of the harvested energy \emph{quanta} in a cycle can be written as
\begin{equation}
 p_E(k) =
 \begin{cases}
   F_E\left((k+1)\Etx\right) - F_E\left(k\Etx\right) & \text{ for } 0\leq k < N_s \\
   1 - F_E(N_s\Etx) & \text{ for } k = N_s,
 \end{cases}
\label{quapdf}
\end{equation}
where $F_E(\cdot)$ is the continuous CDF of the harvested energy $E_h$ defined in (\ref{ecdf}). Note that the domain of $p_E(\cdot)$ is limited between $0$ and $N_s$, thus taking into account the finite battery size $G$.

The number of energy quanta that can be spent during the cycle is instead equal to the number of time slots in the TP (capped to $N_s$), and proportional to the inter vehicle distance, which is distributed according to $f_D(\cdot)$. Henceforth, it follows the distribution
\begin{equation}
 p_T(k) = \begin{cases}
           F_D(2\ell) & \text{ for } k = 0, \\
           F_D(2\ell + kv_0T) + & \\
           \,\,- F_D(2\ell + (k-1)v_0T) & \text{ for } 0 < k < N_s, \\
           1 - F_D(2\ell + (N_s-1)v_0T) & \text{ for } k = N_s.
          \end{cases}
\label{txdist}
\end{equation}
In fact, when the inter vehicle distance is lower than $2\ell$, the cycle has no TP, no transmission is performed, and no energy is depleted. Conversely, if the  distance is greater than $2\ell+(N_s-1)v_0T$, the maximum number $N_s$ of transmission is limited by the battery capacity.

The probability $p_{i,j}$ of the process moving from state $s_q=i\Etx$ to state $s_{q+1}=j\Etx$ can be computed as
\begin{equation}
 p_{i,j} \!= \begin{cases}
           \displaystyle \sum_{h=\max(j-i,0)}^{N_s}\!\!\!\!\!\!\!p_E(h)p_T\left(\min(i+h,N_s)-j\right) & \text{ if } j\neq 0 \\
           \displaystyle \sum_{h=0}^{N_s}p_E(h)\sum_{k=\min(i+h,N_s)}^{N_s}p_T(k) & \text{ if } j = 0.
          \end{cases}
\end{equation}
These values, for $i,j,\in\{0,1,2,\ldots,N_s\}$ are collected in the $N_s+1\times N_s+1$ transition matrix $\mathbf{M}$, with element $\mathbf{M}[i,j]$ equal to $p_{i-1,j-1}$. The steady state distribution $\mathbf{\pi}$ is then obtained as the normalized left eigenvector of $\mathbf{M}$ associated with eigenvalue 1.

We point out a couple of observations.
Firstly, the distribution $p_E(\cdot)$ assumes that energy is harvested from one vehicle at a time. This may not be entirely true if two vehicles are very close to each other, and especially if their distance is lower than $2\ell$.
In this case, the proposed theoretical approach is an approximation that tends to underestimate EH, whose tightness depends on the value of $\ell$ and on the inter-vehicle distance distribution $f_D(\cdot)$. As long as the energy harvested from a vehicle when its distance is higher than $\ell$ is negligible, and/or $F_D(2\ell)$ is low, the theoretical result is expected to faithfully reproduce the actual system behavior.
Secondly, the derivation of the distribution $p_B(\cdot)$ of the battery status presented above makes it hard to derive a closed form expression to the maximization problem (\ref{optsim}). Nevertheless, it offers a simple yet effective tool to analyze the throughput as a function of several tunable parameters, and to draw some practical conclusions, as detailed in the Results section.

\section{Throughput computation}
\label{sec:throcomp}
Exploiting the battery status quantization, the expression in (\ref{optint}) can be rewritten as
\begin{equation}
 \Thr = \frac{\phi_sv_0}{\mathbb{E}[d_v]}\sum_{k=0}^{N_s}p_B(k)\Psi(k),
 \label{genthro}
\end{equation}
where
\begin{eqnarray}
 \Psi(k) \!\!& = \!\!& \int_{2\ell}^{+\infty}\int_0^{+\infty}\min\left(\frac{y-2\ell}{v_0T},\frac{\min(k\Etx+x,G)}{\Etx}\right)\times\nonumber \\
 & & \times\,\,f_E(x)f_D(y)\de x\de y
 \label{newthro}
\end{eqnarray}
is the average number of transmissions per cycle, for a given value of the tunable parameter $\ell$, conditioned on the fact that the battery contains $k$ energy quanta at the beginning of the cycle. In (\ref{newthro}) we have accounted for the fact that throughput is 0 when the inter vehicle distance $y$ is lower than $2\ell$, since in this case there is no TP.

In order to solve (\ref{newthro}), we can firstly split the inner integral into two terms. Recalling that $G=N_s\Etx$, we get
\begin{eqnarray}
 \Psi(k) \!\!\!\!\!& =\!\!\!\!\! & \int_{2\ell}^{+\infty}\!\!\!\int_0^{\delta}\!\!\min\left(\frac{y-2\ell}{v_0T},\frac{x}{\Etx}+k\right)\!f_E(x)\de x f_D(y)\de y + \nonumber \\
  & &\!\!\!\! + \left(1-F_E(\delta)\right)\int_{2\ell}^{+\infty}\!\!\min\left(\frac{y-2\ell}{v_0T}, N_s\right)f_D(y)\de y,
  \label{splitthro}
\end{eqnarray}
being $\delta = (N_s-k)\Etx$ the amount of energy needed to fully replenish the battery. The latter term, which we call $\Psi_2(k)$, represents the case when the harvested energy recharges the battery completely. The number of transmissions here depends only on the battery capacity and the cycle duration (that is, the distance of the next vehicle), and therefore only on $f_D(\cdot)$. We can elaborate it to get
\begin{eqnarray}
 \Psi_2(k) & = & \frac{1-F_E(\delta)}{v_0T}\left(\int_{2\ell}^Wyf_D(y)\de y + N_sv_0T +\right. \nonumber\\
 & &  + 2\ell F_D(2\ell) - WF_D(W)\Bigg), \nonumber \\
 & = & \frac{1-F_E(\delta)}{v_0T}\left(N_sv_0T -\int_{2\ell}^WF_D(y)\de y\right)\!\!,
\end{eqnarray}
where $W = 2\ell+N_sv_0T$ is the maximum inter-vehicle distance over which a fully charged battery can avoid an energy outage.

The former term in (\ref{splitthro}), which we call $\Psi_1(k)$, corresponds to the case when the battery is not fully replenished during the harvest phase of the current cycle. In this case, we can split the outer integral into three parts, each corresponding to a different situation:
\begin{enumerate}[leftmargin=*] 
 \item $2\ell < y \leq 2\ell + kv_0T = R$: in this case, the inter vehicle distance is low, and the energy contained in the battery even before the harvest phase ($k\Etx$) is already enough to perform all the transmissions of the cycle; the corresponding term is
 \begin{eqnarray}
  \Psi_{11}(k) & = & \int_{2\ell}^R\int_0^{\delta}\frac{y-2\ell}{v_0T}f_E(x)\de x f_D(y)\de y \nonumber \\
  & \hspace{-2.5cm} = & \hspace{-1.5cm}\frac{F_E(\delta)}{v_0T}\left(\int_{2\ell}^Ryf_D(y) \de y - 2\ell\left(F_D(R) - F_D(2\ell)\right)\right)\nonumber \\
  & \hspace{-2.5cm} = & \hspace{-1.5cm}\frac{F_E(\delta)}{v_0T}\left(kv_0TF_D(R) - \int_{2\ell}^RF_D(y) \de y\right).
 \end{eqnarray}

 \item $R < y \leq W$: in this case, the inter vehicle distance is greater, the energy already stored in the battery is not sufficient, and the overall number of transmissions depends on both the amount of harvested energy and the cycle length; the corresponding term is
 \begin{eqnarray}
  \Psi_{12}(k) &\!\!\!\!\! = &\!\!\!\!\! \int_R^W\!\!\!\int_0^{\delta}\left[k+\frac{\min\left(\beta(y),x\right)}{\Etx}\right]f_E(x)\de x f_D(y)\de y \nonumber \\
  &\hspace{-2.2cm} = &\hspace{-1.1cm}\!\! \frac{1}{\Etx}\!\!\int_R^W\!\!\!\left(\!\int_0^{\beta(y)}\!\!\!\!\!\!\!\!xf_E(x)\de x \!+\! \beta(y)\int_{\beta(y)}^{\delta}\!\!\!\!\!\!f_E(x)\de x\!\right)\!\!f_D(y)\de y\! +\nonumber\\
  &\hspace{-2.2cm} &\hspace{-1.1cm}\!\! + kF_E(\delta)\left(F_D(W)-F_D(R)\right),
  \label{psi12}
 \end{eqnarray}
 where $\beta(y) = [(y-2\ell)/(v_0T) - k]\Etx$ is the minimum energy to be harvested to avoid energy outage, and is a function of $y$. After some algebraic manipulations, $\Psi_{12}(k)$ can be rewritten as
 \begin{eqnarray}
  \Psi_{12}(k) &\hspace{-.3cm} = & \hspace{-.3cm} F_D(W)F_E(\delta)N_s + \frac{1-F_D(W)}{\Etx}\!\int_0^\delta\!\! F_E(y)\de y + \nonumber \\
  &\hspace{-.7cm} &\hspace{-.7cm} -\frac{F_E(\delta)}{v_0T}\int_R^W F_D(y)\de y - kF_E(\delta)F_D(R) + \nonumber\\
  &\hspace{-.7cm} &\hspace{-.7cm} -\frac{1}{\Etx}\int_0^\delta F_E(y)\left(1-F_D\left(\frac{v_0T}{\Etx}y+R\right)\right)\de y.
 \end{eqnarray}

 \item $y > W$: in this case, the inter vehicle distance is large, and even replenishing the battery is not enough to guarantee continuous transmissions until the end of the current cycle. The overall number of transmissions depends only on the amount of harvested energy, and the corresponding term is
 \begin{eqnarray}
  \Psi_{13}(k) &\!\!\!\! = &\!\!\!\! \int_W^{+\infty}\int_0^{\delta}\left(k+\frac{x}{\Etx}\right)f_E(x)\de x f_D(y)\de y \nonumber \\
  &\hspace{-2.4cm} = &\hspace{-1.4cm} \left(1-F_D(W)\right)\left(k F_E(\delta) + \frac{1}{\Etx}\int_0^{\delta}\!\!xf_E(x)\de x\right)\nonumber\\
  &\hspace{-2.4cm} = &\hspace{-1.4cm}\left(1-F_D(W)\right)\left(N_s F_E(\delta)-\frac{1}{\Etx}\int_0^{\delta}\!\!F_E(x)\de x\right)\!\!.
 \end{eqnarray}
\end{enumerate}

Overall, the expected number of transmissions per cycle, conditioned on the battery level $k$, is given by $\Psi_{11}(k) + \Psi_{12}(k)+\Psi_{13}(k)+\Psi_2(k)$, which yields
\begin{eqnarray}
 \Psi(k) &\!\!\!\! = &\!\!\!\! N_s - \frac{1}{v_0T}\int_{2\ell}^WF_D(y)\de y + \nonumber\\
 &\!\!\!\! &\!\!\!\! - \frac{1}{\Etx}\int_0^\delta \!\!F_E(y)\left[1-F_D\left(\frac{y}{m}+R\right)\right] \de y,
 \label{evothro}
\end{eqnarray}
with $m=P_t/v_0$.
Intuitively, equation (\ref{evothro}) states that the maximum number of transmitted packets in a cycle would be $N_s$, that is, the number of packets that can be sent with a fully charged battery. This number is limited by two factors: \emph{i)} the subsequent vehicle can be too close to allow the EHD to send all the packets (second term in (\ref{evothro})); \emph{ii)} the energy harvested may not be enough to charge the battery and sustain continuous transmission when the subsequent vehicle is farther away (third term in (\ref{evothro})).

\subsection{Sparse vehicular traffic: Poisson arrivals}
In this section, we consider a scenario with low vehicular traffic, which we model as a Poisson process, with intensity $\mu$, meaning that $F_D(x) = 1 - e^{-\mu x}$. This distribution is not fully realistic, since its domain is $\mathbb{R}^+$, while very short distances are not feasible in real scenarios. However, as long as $\mu$ is not too high, it can still give reasonable results. In addition, slight modifications, likely introducing a minimum distance between vehicles, can be easily incorporated.
Under this assumption, substituting $F_D(x)$ it into (\ref{evothro}) gives
\begin{eqnarray}
 \Psi(k) & = & N_s - \frac{1}{v_0T}\left(W-2\ell+\frac{e^{-\mu W}}{\mu} - \frac{e^{-2\ell\mu}}{\mu}\right) + \nonumber \\
 & & - \frac{e^{-\mu R}}{\Etx}\int_0^\delta F_E(y)e^{-\mu \frac{y}{m}} \de y \nonumber \\
 &\hspace{-1cm} = &\hspace{-.8cm}  \frac{1}{\mu v_0T}\left(e^{-2\ell\mu}\! - e^{-\mu R}\right)\! +\! \frac{e^{-\mu R}}{\Etx}\!\!\int_0^\delta\!\! (1-F_E(y))e^{-\mu \frac{y}{m}} \de y. \nonumber\\
 \label{poithro}
 \end{eqnarray}
The integral, using the CDF in (\ref{ecdf2}), can be numerically computed. By replacing (\ref{poithro}) into (\ref{genthro}), and setting $\mathbb{E}[d_v]=1/\mu$, the average throughput $\Thr$ is obtained.

\subsection{Heavy vehicular traffic or platooning: constant inter-vehicle distance}
Due to the foreseen spread of V2X communications and autonomous vehicles, platoons are expected to be often exploited as an energy saving, safe formation. Vehicles in a platoon move at the same speed, and keeping inter-vehicle distances approximately constant.
We model this scenario by replacing the stochastic inter-vehicle distance $d_v$ with a fixed one, namely $\dpla$. A very similar situation is represented by vehicular traffic congestion, where vehicles move (usually at low speed) with nearly constant distance from each other.

In this scenario, the distribution $p_T(k)$ of the energy quanta depleted, in (\ref{txdist}), becomes equal to 1 only for $k=\lfloor \min((\dpla-2\ell),N_s)/(v_0T)\rfloor$, and 0 otherwise.
The expression for the conditioned number $\Psi(k)$ of transmissions per cycle greatly simplifies, and depends on the value $\dpla$, thus giving 
\begin{equation}
 \Psi(k) = \left\{
 \begin{array}{lr}
  0 & \hspace{-1.8cm}\text{if } \dpla < 2\ell \\
  \displaystyle\frac{1}{v_0T}\left(\dpla-2\ell\right) & \hspace{-1.8cm}\text{if } 2\ell \leq \dpla < R \\
  \displaystyle k + \frac{1}{\Etx}\int_0^{m(d_0-R)}\!\!\!\!\!\!\!\!\!\!\!\!(1-F_E(x))\de x & \text{if } R\leq \dpla < W \\
  \displaystyle k + \frac{1}{\Etx}\int_0^\delta(1-F_E(x))\de x & \text{if } \dpla \geq W,
 \end{array}
 \right.
 \label{throplat}
\end{equation}
where the integral, inserting the CDF in (\ref{ecdf2}), can be easily computed numerically.
By putting (\ref{throplat}) into (\ref{genthro}), with $\mathbb{E}[d_v] = \dpla$, the expected throughput is obtained. 
Notice that in this case, since the duration of the cycle is constant, the expression in (\ref{genthro}) does not require the Renewal-Reward process approximation, and is therefore exact.
In this scenario, $\dpla$ depends on the vehicular traffic conditions. Nevertheless, it can still be partially tuned by the EHD, by choosing not to harvest energy from every vehicle, but, if convenient, from only some of them, at regular intervals, which is equivalent to have a bigger inter-vehicle distance ($2\dpla$, $3\dpla$, and so on).

\subsection{Energy Efficiency}
We can use the throughput expression to derive energy efficiency, which is also a relevant metric in this scenario.
We compute it, in both the abovementioned scenarios, as the expected number of bits that are correctly delivered per unit of available energy.

For the platooning scenario, with inter-vehicle distance $\dpla$, consider a cartesian plane, where vehicles move along the $x$-axis heading to the left.
The EHD is placed aside from the street, at location $(0,-w)$.
For symmetry reasons, we can evaluate the average RF energy density $\Epla$ over the time interval $T_c = [-\dpla/(2v_0),\dpla/(2v_0)]$, where we assume, without loss of generality, that at instant $0$ the closest vehicle, $V_0$, is at location $(0,0)$. 

The energy density in this time interval can be written as
\begin{equation}
 \Epla = \frac{v_0}{\dpla}\int_{-\frac{\dpla}{2v_0}}^{\frac{\dpla}{2v_0}}\sum_j P_r^{(j)}(t)\de t,
 \label{avaene}
\end{equation}
being $P_r^{(j)}(t)$ the power received from the $j$--th vehicle.
Equation (\ref{avaene}) is exact if the signals received from different sources are nonoverlapping, narrow-band signals~\cite{VEH17}, which is reasonable at least from neighboring vehicles, in order to avoid interference. Indeed, in practical cases the energy coming from far vehicles is negligible, and the summation in (\ref{avaene}) can be considered a very good approximation.

We compute $P_r^{(j)}(t)$ by averaging the effect of the fast fading, thus getting:
\begin{equation}
 P_r^{(j)}(t) = \frac{P_v}{\left(w^2+(v_0t-j\dpla)^2\right)^{\alpha/2}},
\end{equation}
where we consider the path loss attenuation.

Equation (\ref{avaene}) can be hence reformulated as
\begin{eqnarray}
 \Epla & = & \frac{P_vv_0}{\dpla}\sum_{j=-\infty}^{+\infty}\int_{-\frac{\dpla}{2v_0}}^{\frac{\dpla}{2v_0}}\frac{1}{\left(w^2+(v_0t-j\dpla)^2\right)^{\alpha/2}}\de t \nonumber \\
 & = & \frac{P_v}{\dpla}\sum_{j=-\infty}^{+\infty}\int_{-\frac{\dpla}{2}}^{\frac{\dpla}{2}}\frac{1}{\left(w^2+(x-j\dpla)^2\right)^{\alpha/2}}\de x,
\end{eqnarray}
where we set $x=v_0t$, thus converting time into space. Now, by further setting $y=x-j\dpla$, we obtain
\begin{eqnarray}
 \Epla & = & \frac{P_v}{\dpla}\sum_{j=-\infty}^{+\infty}\int_{-j\dpla-\frac{\dpla}{2}}^{-j\dpla+\frac{\dpla}{2}}\frac{1}{\left(w^2+y^2\right)^{\alpha/2}}\de y \nonumber \\
 & = & \frac{P_v}{\dpla}\int_{-\infty}^{+\infty}\frac{1}{\left(w^2+y^2\right)^{\alpha/2}}\de y \nonumber\\
 & = & \frac{P_v}{\dpla}\frac{\sqrt{\pi}w^{1-\alpha}\Gamma\left(\frac{\alpha-1}{2}\right)}{\Gamma\left(\frac{\alpha}{2}\right)},
 \label{genene}
\end{eqnarray}
where $\Gamma(\cdot)$ is the Gamma function.
Notice that the average amount of energy available during $T_c$ is equal to the one that would be provided over an infinite time horizon by vehicle $V_0$ alone. This is reasonable, since the energy that $V_0$ provides out of $T_c$ must be equal to the energy that other vehicles provide within $T_c$, due to the symmetry of the configuration.

The same expression holds also for the scenario with Poisson arrivals. In a generic instant $t_0$, call $x_i$ the position, over the $x$ axis, of the $i$--th vehicle. The average energy density here reads as
\begin{eqnarray}
 \Epoi & = & \mathbb{E}\left[\sum_i\frac{P_v}{(w^2+x_i^2)^{\alpha/2}}\right] \nonumber\\
 & \stackrel{\text{(a)}}{=} & \mu P_v\int_{\mathbb{R}}\frac{1}{(w^2+x^2)^{\alpha/2}}\de x,
\end{eqnarray}
where (a) comes from Campbell's Theorem applied to a linear Poisson Point Process with intensity $\mu$, and gives back (\ref{genene}), with $\dpla$ replaced by $1/\mu$. Since these two quantities represent the expected inter vehicle distance in the two scenarios, respectively, we can replace them both with $\mathbb{E}[d_v]$, thus getting a unique expression which, for the case $\alpha=3$, can be simplified into
\begin{equation}
 \varepsilon = \frac{2 P_v}{w^2\mathbb{E}[d_v]}.
\end{equation}

The throughput, in bits per second, is $S\Thr$, being $S$ the packet size. The overall energy efficiency, that is, the number of bits correctly sent per unit of available energy, is therefore
\begin{equation}
 \Upsilon = \frac{S\Thr}{\varepsilon} = \frac{Sw^2\mathbb{E}[d_v]}{2 P_v}\Thr.
\end{equation}

\section{Results}
\label{sec:resu}
We analyze how different parameters, like the vehicular traffic pattern and intensity, the battery capacity, the EHD transmission power and the choice of the harvesting distance affect the system throughput. In addition, we compare the theoretical results with those obtained through extensive MATLAB simulations.
The parametes setup is similar to \cite{VEH3}, and unless otherwise specified, their values are those reported in Table~\ref{tab:values}, while the Rice factor is set to $\kappa=10$ dB.
For the sake of clarity, in all the figures the throughput is reported in kbit/s, and is obtained by multiplying $\Thr$, which is measured in pkt/s, by the packet size $S$.

\begin{table}
 \centering
 \begin{tabular}{cc|cc}
  \hline\hline
  Parameter & Value & Parameter & Value\\
  \hline $r$ & 200 m & $N_0$ & -90 dBm\\
  $w$ & 5 m & $\alpha$ & 3\\
  T & 100 ms& $\eta$ & 0.5\\
  $v_0$ & 10 m/s & $B$ & 15 kHz\\
  $P_t$ & 40 $\mu$W & $S$ & 1 kbit\\
  $P_v$ & 100 mW & $G$ & 400 $\mu$J\\
  \hline\hline\\
 \end{tabular}
 \caption{Parameter values.}
 \label{tab:values}
 \vspace{-0.8cm}
\end{table}

Our proposed cycle-based strategy requires that the EHD has local topology knowledge (with radius slightly higher than the selected harvest distance $\ell$) and of the vehicular traffic type (inter-vehicle distance statistics) to properly set its tunable parameters.

\subsection{Analysis validation}
\begin{figure}
 \begin{center}
   \includegraphics[width=\figww]{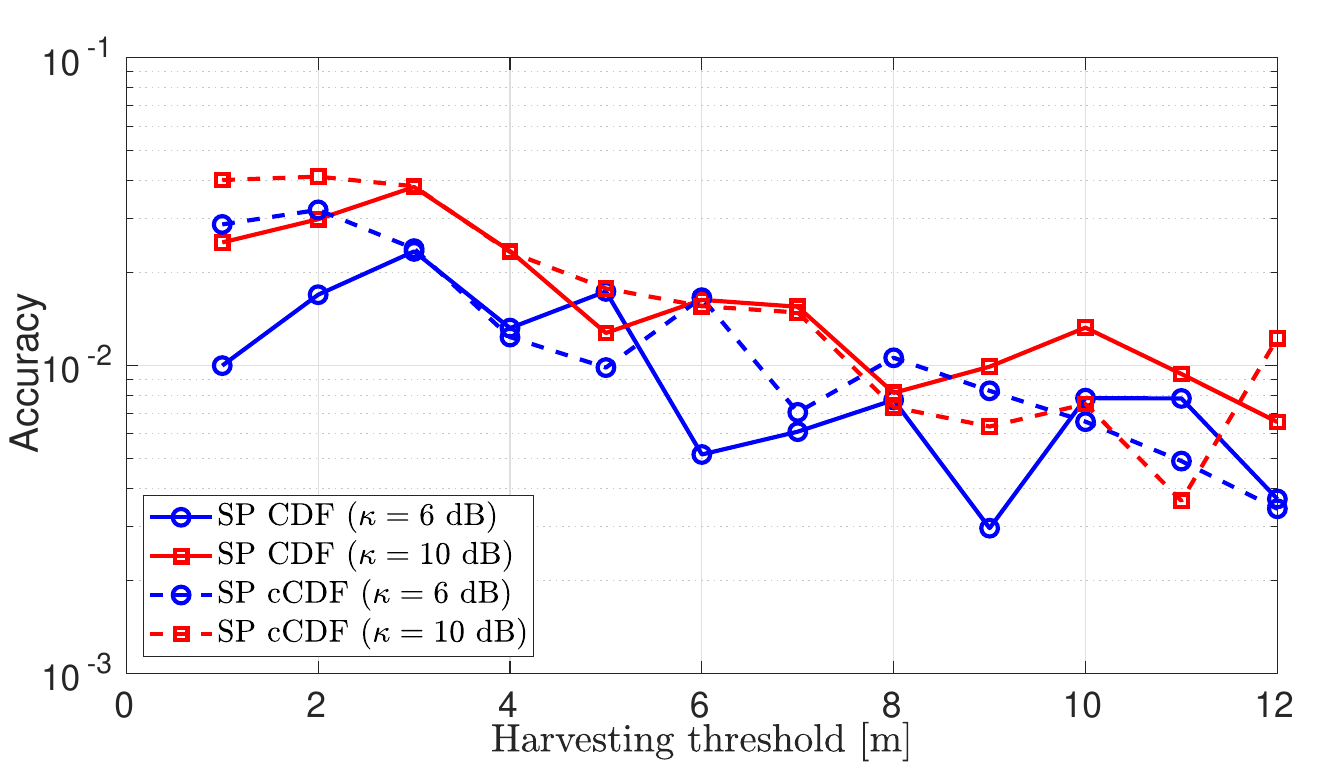}
 \caption{\small Accuracy of the CDF and cCDF of the harvested energy obtained via saddle point approximation.}
 \label{fig:AccuSP}
 \vspace{-0.5cm}
 \end{center}
\end{figure}

In order to assess the validity of the proposed theoretical approach, we first check the tightness of the saddle point approximation for the CDF $F_E(\cdot)$ of the harvested energy derived in (\ref{ecdf2}). We use the accuracy metric defined in~\cite{approsum} to measure the relative difference between the theoretical CDF and the real one, $\hat F_E(\cdot)$, obtained through Monte Carlo simulations with $10^6$ runs. The range of interest is the interval where $0.001<\hat F_E(\cdot)<0.5$. The accuracy is also derived for the complementary CDF (cCDF); in this case, we choose as range of interest the interval where $0.5<\hat F_E(\cdot)<0.999$. The obtained values are reported in Figure~\ref{fig:AccuSP}, as a function of the harvesting threshold $\ell$. The accuracy is always below 0.04, and decreases down to 0.003, for $\kappa=6$ dB, as the number $L$ of summands increases, thus validating the approximation tightness.

\begin{figure}
 \begin{center}
   \includegraphics[width=\figww]{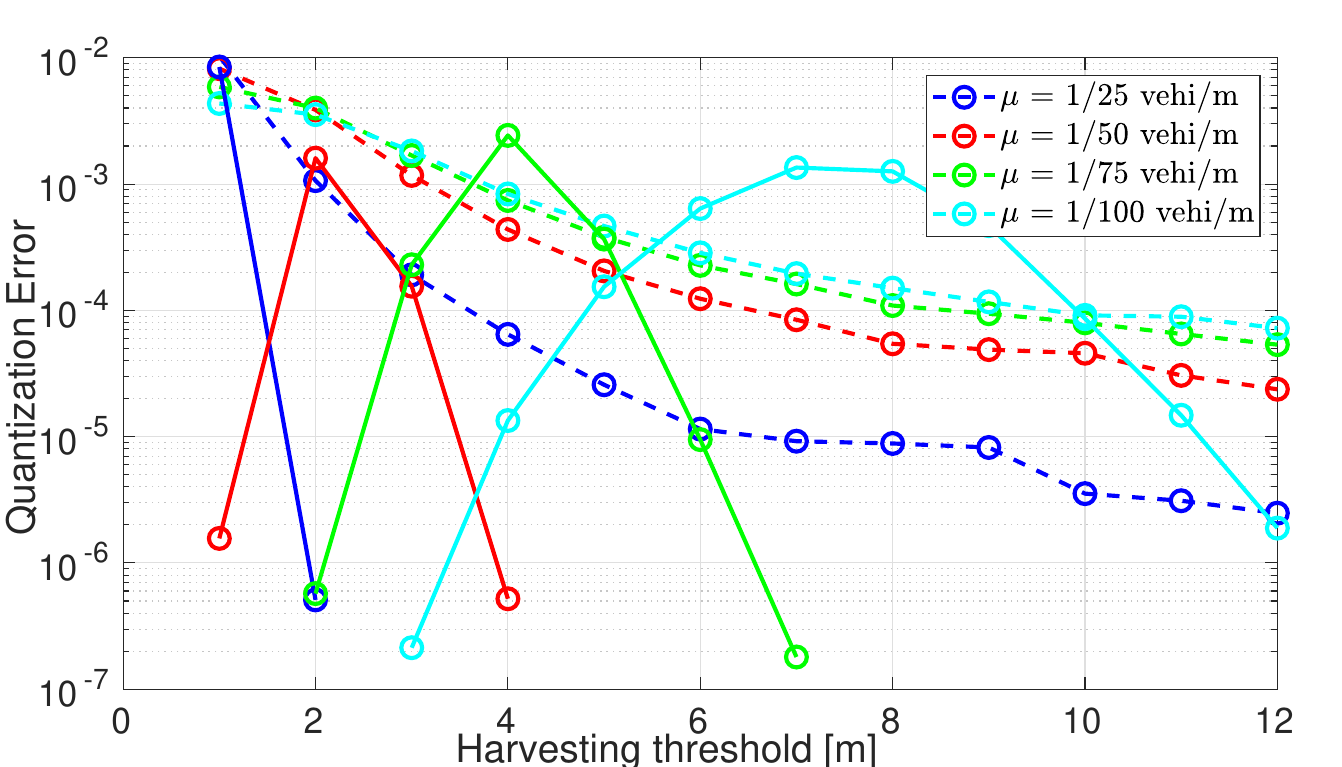}
 \caption{\small Quantization error as a function of the harvesting threshold. Solid lines are for the platooning scenario, while dashed lines are for the sparse vehicular traffic scenario.}
 \label{fig:QuantErr}
 \vspace{-0.5cm}
 \end{center}
\end{figure}

Another approximation comes from the energy quantization adopted in the theoretical model in equation (\ref{quapdf}). We assessed the throughput error due to this quantization through extensive simulations, comparing the scenarios with and without energy quantization.
The obtained results are plotted in Figure~\ref{fig:QuantErr}. For the platooning scenario and a given $\mu$, the error is lower when $\ell$ is either very low, meaning that the battery is emptied at almost every cycle, or very high, causing frequent energy overflows. Both events in fact lead to the same battery status (empty or full), irrespective of the quantization. In the sparse vehicular traffic scenario, instead, the variance in the inter-vehicle distance leads to a more smooth behavior of the error curve, decreasing with $\ell$. In all cases, however, the quantization error is always lower than 1\%

\begin{figure}
 \begin{center}
   \includegraphics[width=\figww]{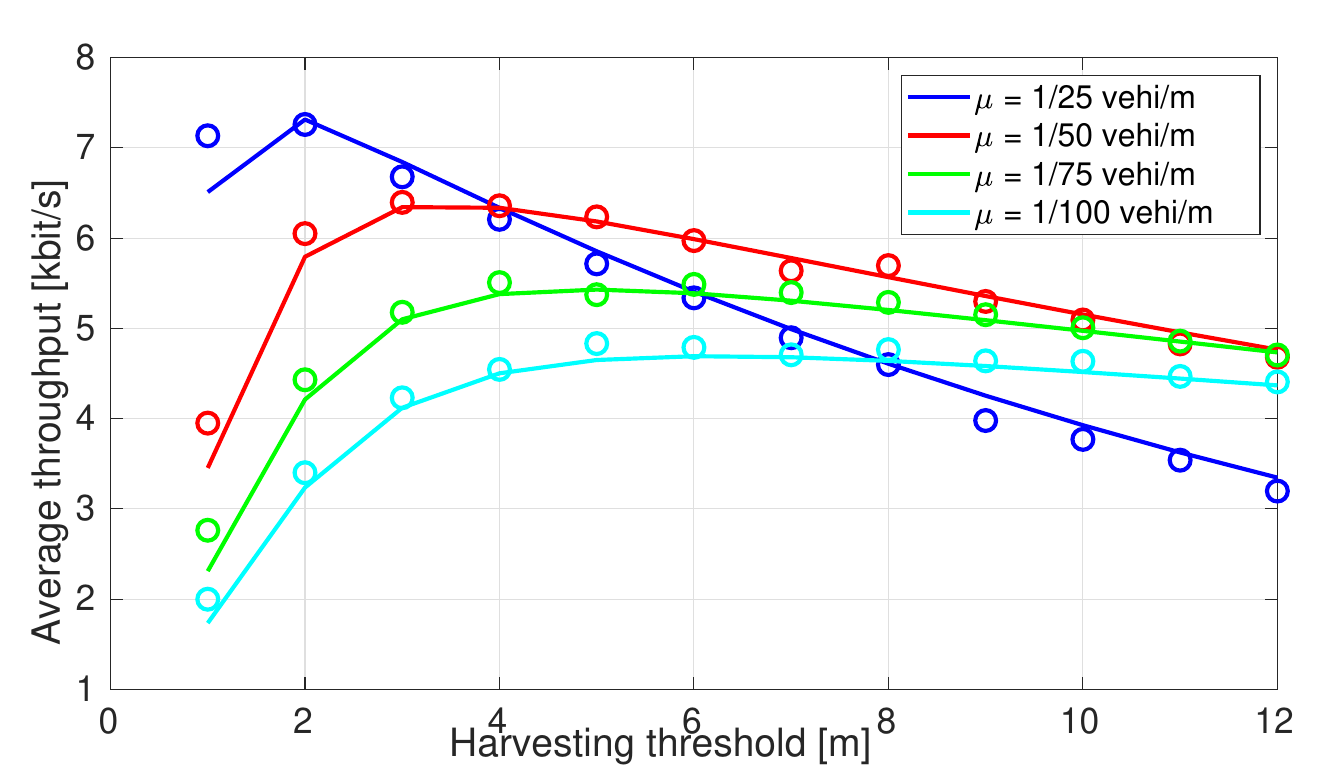}
 \caption{\small Average throughput as a function of the harvesting distance $\ell$, with Poisson arrivals. Lines are theoretical results, while markers are simulation results.}
 \label{fig:varlim_Poi}
 \vspace{-0.5cm}
 \end{center}
\end{figure}
In Figure~\ref{fig:varlim_Poi}, we investigate the performance of the proposed strategy in the low vehicular traffic scenario, with arrivals modeled through a Poisson process. We vary $\ell$ with step size 1 m (smaller step sizes might exceed the localization precision of the vehicles).
The first observation is that the throughput expression based on (\ref{poithro}) matches the simulation results for all the parameters settings, thus confirming the validity of the analysis.
The minor discrepancy observed for low values of $\ell$ is due to the fact that in this case the energy harvested from other vehicles may become non negligible, and the theoretical results slighlty underestimate the real throughput.
Secondly, we notice that almost all curves show a unique maximum. As expected, lowering $\ell$ too much causes frequent energy outage events, thus limiting the number of transmissions; increasing it too much reserves too much time for energy harvesting, hence limiting the throughput.
Thirdly, the vehicular traffic intensity has an impact. When more vehicles are present, energy is more abundant, and it is preferable to reduce $\ell$, in order to leave more time for transmissions. Indeed, using high values for $\ell$ with high vehicular traffic intensities (e.g., $\mu=1/25$ vehicles/m) is particularly detrimental, since several TPs may have very short or even null duration: the overall number of transmissions is hence reduced, and the high amount of energy harvested in long HPs is mostly wasted due to battery overflow. Conversely, when the traffic is less congested, the optimal value of $\ell$ increases (doubling from $\mu=1/50$ vehicles/m to $\mu=1/100$ vehicles/m). 

\begin{figure}
 \begin{center}
   \includegraphics[width=\figww]{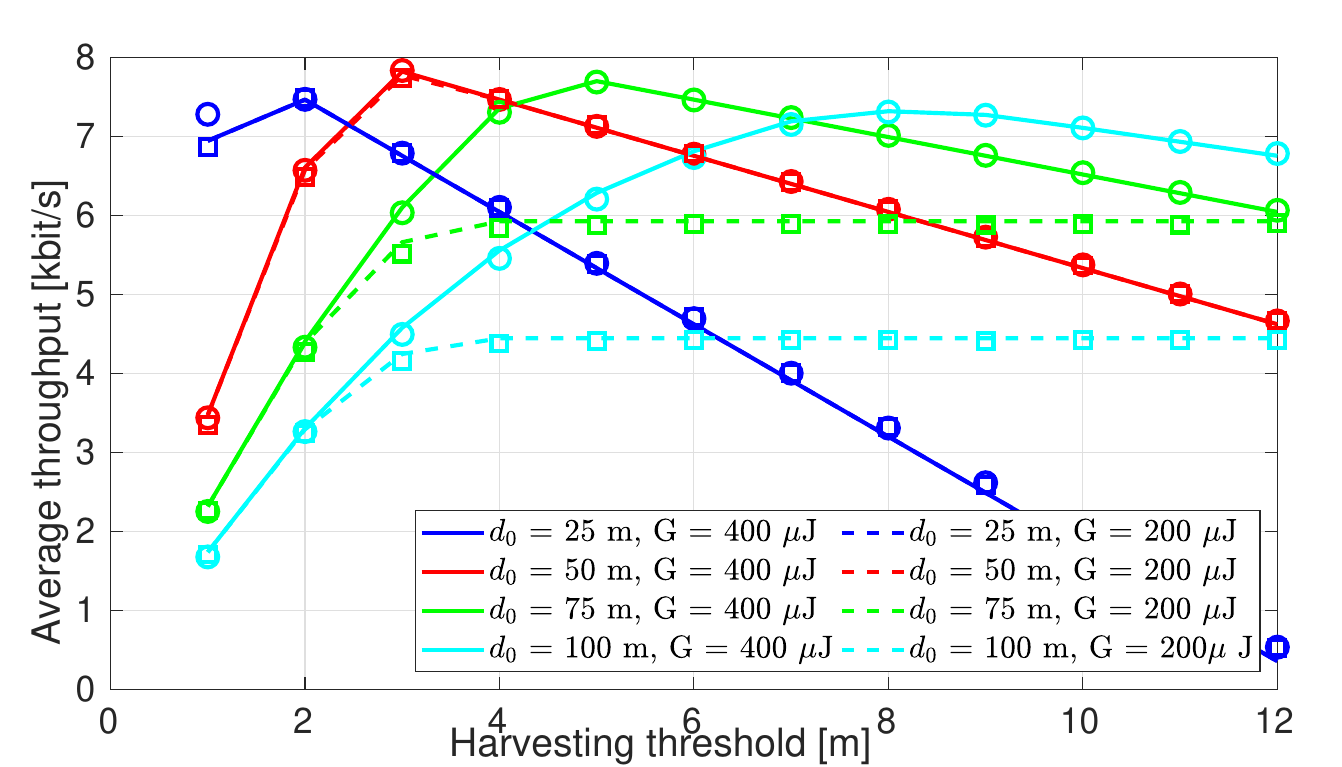}
 \caption{\small Average throughput as a function of the harvesting distance $\ell$, with fixed inter-vehicle distance. Lines are theoretical results, while markers are simulation results.}
 \label{fig:varlim_Fix}
 \vspace{-0.2cm}
 \end{center}
 \vspace{-.5cm}
\end{figure}
Figure~\ref{fig:varlim_Fix} depicts the results with the same parameters, in the heavy vehicular traffic scenario, where the inter vehicle distance is approximately fixed. Notice that the value $\dpla$ does not necessarily correspond to the distance between two adjacent vehicles, since the EHD might choose to harvest only from a subset of the vehicles. Indeed, this appears to be a smart choice, since the maximum achievable throughput does not monotonically grow as $\dpla$ decreases, as it does in the low vehicular traffic scenario.
This is mainly due to the limited battery size, which makes the choice of harvesting too often less advantageous.
We also observe that the achievable throughput in the platooning scenario is higher than that in low traffic scenario, even when the average inter vehicle distance is higher. For instance, the maximum throughput attained in a Poisson modeled vehicular traffic with $\mu = 1/50$, with average inter vehicle distance equal to $50$ m, is between 6.2 and 6.5 kbit/s; the maximum throughput in the platooning scenario with $\dpla=100$ m is instead higher than 7 kbit/s.
The presence of regularly spaced energy sources is beneficial, since it avoids long periods of energy outage which can occur in the sparse vehicular traffic scenario. Furthermore, battery overflow events due to very close vehicles are also avoided. We can state that harvesting energy from V2V communications is more convenient when vehicles are more regularly spaced.

In the same figure, we also plot the lines for the case of a halved battery capacity (200 $\mu$J). The limited battery size has a pronounced effect on the curves for higher $\ell$, since these values correspond to more harvested energy, which is likely to be wasted due to energy overflow.
The throughput reaches a plateau when $\ell$ becomes high enough to allow full battery replenishment at the end of the HP.

\subsection{Impact of transmit power, data rate and vehicular traffic intensity}
Apart from the harvest distance $\ell$, the transmission power $P_t$ has also a strong impact on the overall throughput, especially if the transmission rate grows. On one side, transmitting at higher power increases the delivery probability; on the flip side, however, energy is depleted more rapidly, thus increasing the frequency of energy outage events.
\begin{figure}
 \begin{center}
   \includegraphics[width=\figww]{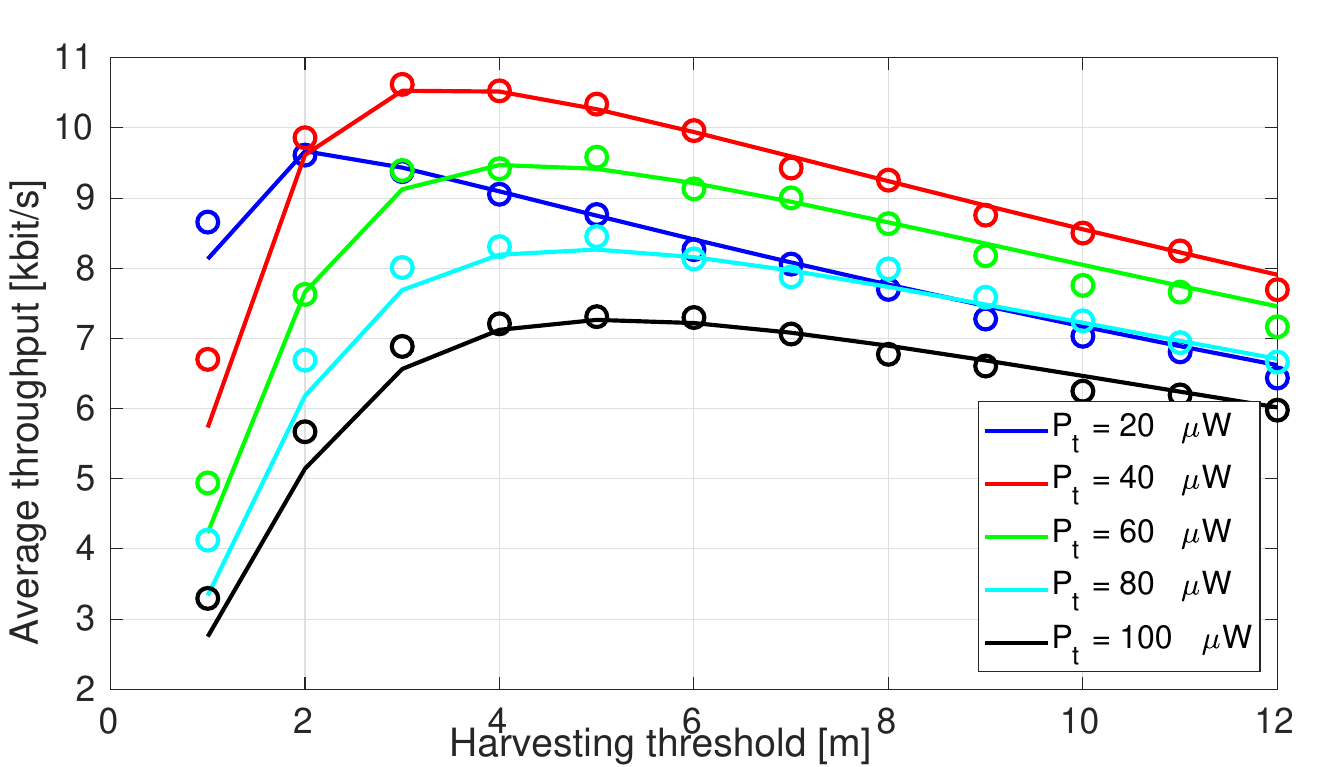}
 \caption{\small Average throughput for varying values of the transmit power $P_t$, low vehicular traffic scenario with $\mu = 1/50$ vehicles/m. Lines are theoretical results, while markers are simulation results.}
 \label{fig:varlim_Pt_Poi}
 \vspace{-0.5cm}
 \end{center}
\end{figure}
\begin{figure}
 \begin{center}
   \includegraphics[width=\figww]{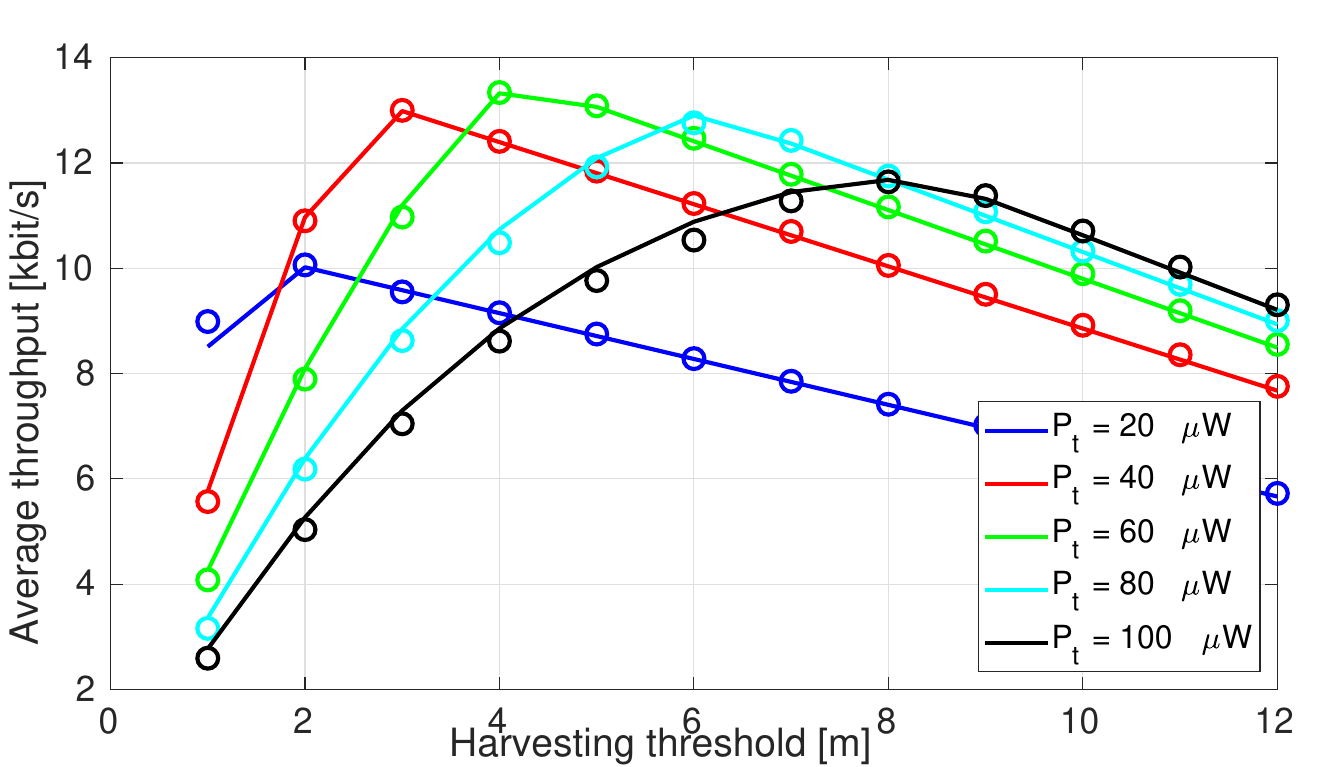}
 \caption{\small Average throughput for varying values of the transmit power $P_t$, high vehicular traffic scenario with $\dpla = 50$ m. Lines are theoretical results, while markers are simulation results.}
 \label{fig:varlim_Pt_Fix}
 \vspace{-0.5cm}
 \end{center}
\end{figure}
We investigate this aspect in Figure \ref{fig:varlim_Pt_Poi}, for the low vehicular traffic scenario, and \ref{fig:varlim_Pt_Fix}, for the high vehicular traffic scenario. We set the packet size to 2000 bit, while the average inter vehicle distance is 50 m in both cases. Five equally spaced power levels are tested.

We observe that transmitting with the lowest power $P_t=20$ $\mu$W is not the best choice, despite granting the lowest amount of energy depletion. Instead, increasing $\ell$ to 4 m and the transmit power to $P_t=40$ $\mu$W in the low traffic scenario grants a 12\% throughput gain with respect to the minimum transmit power level $P_t = 20$ $\mu W$ with $\ell=2$ m, while setting $P_t=60$ $\mu$W in the vehicular high traffic scenario offers an even higher 30\% gain. The reason for the latter scenario allowing a higher value for $P_t$ lies in its bounded inter vehicle distance, which avoids the need to save energy for a potentially very long time period before the arrival of the next vehicle.

\begin{figure}
 \begin{center}
   \includegraphics[width=\figww]{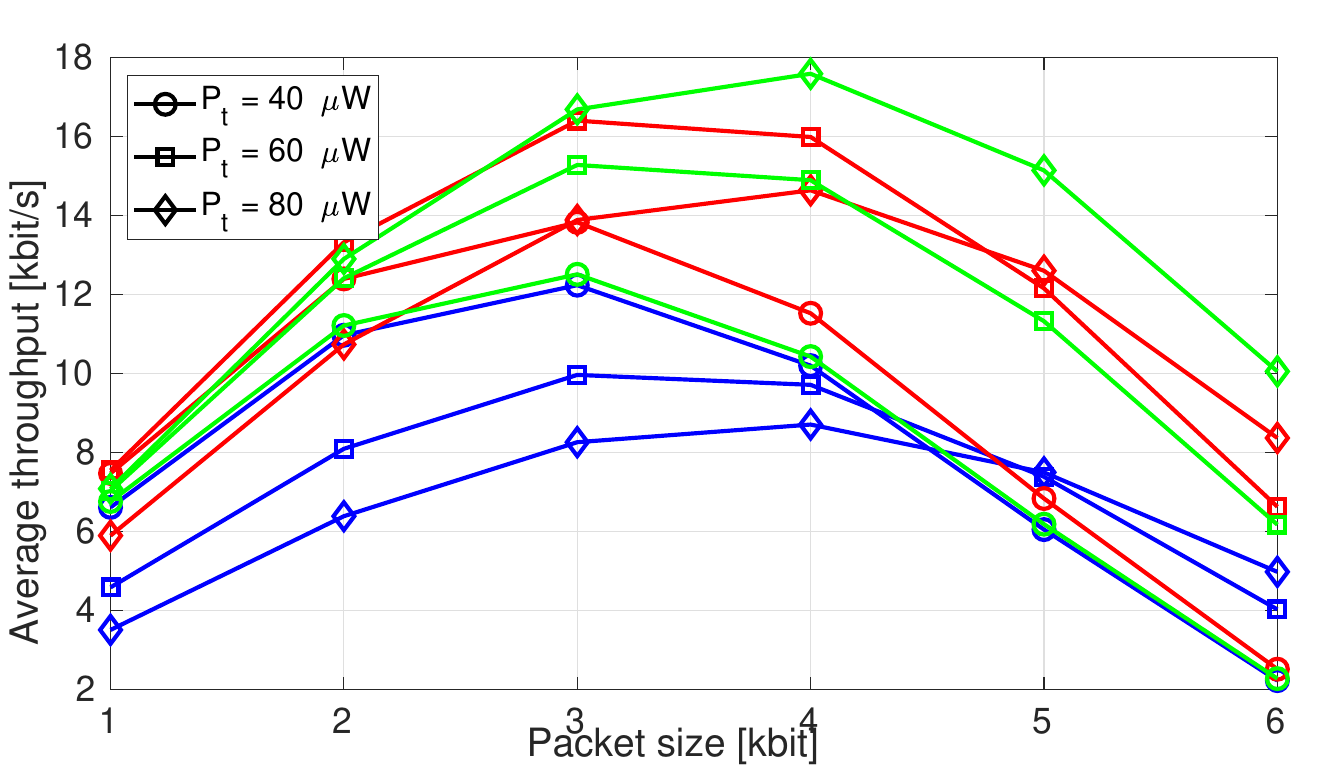}
 \caption{\small Average throughput for varying values of packet size $S$. Blue lines are for $\ell=2$ m, red lines for $\ell = 4$ m and green lines for $\ell = 6$ m. The inter vehicle distance $\dpla$ is set to 50 m.}
 \label{fig:varlen_Pt_Fix}
 \vspace{-0.5cm}
 \end{center}
\end{figure}

The inherent tradeoff in the choice of the transmit power actually depends on the desired data rate, which in our scenario is determined by the packet size $S$, as illustrated in Figure~\ref{fig:varlen_Pt_Fix}, where the high vehicular traffic scenario with $\dpla= 50$ m is analyzed. Since the simulation results tightly match the theoretical ones, in this figure, as well as in the next ones, we do not report them for the sake of legibility.
For low data rate, low transmit power is always preferable, since it still attains a high delivery probability, while limiting the battery energy consumption and hence energy outage. When $S$ grows, however, it is also necessary to increase $P_t$, since the gain in terms of decoding probability at the AP compensates the loss in terms of energy outage probability.
In addition, as $P_t$ grows, the harvest distance $\ell$ also plays a bigger role. For $P_t=40$ $\mu$W, little difference is observed in Figure~\ref{fig:varlen_Pt_Fix} among the lines for the three considered values of $\ell$, and the highest throughput of approximately 14 kbit/s is achieved when $S=3$ kbit with $\ell = 4$ m.
When $P_t = 80$ $\mu$W the maximum throughput (17.68 kbit/s) is obtained at $S=4$ kbit with $\ell = 6$ m, which consistently outperforms the choices of a smaller $\ell$.
This confirms that, as $S$ becomes larger, the maximum throughput can be achieved by increasing both $P_t$, in order to get higher decoding probability at the AP, and $\ell$, in order to limit the energy outage probability, even if this reduces the transmission time.

\begin{figure}
 \begin{center}
   \includegraphics[width=\figww]{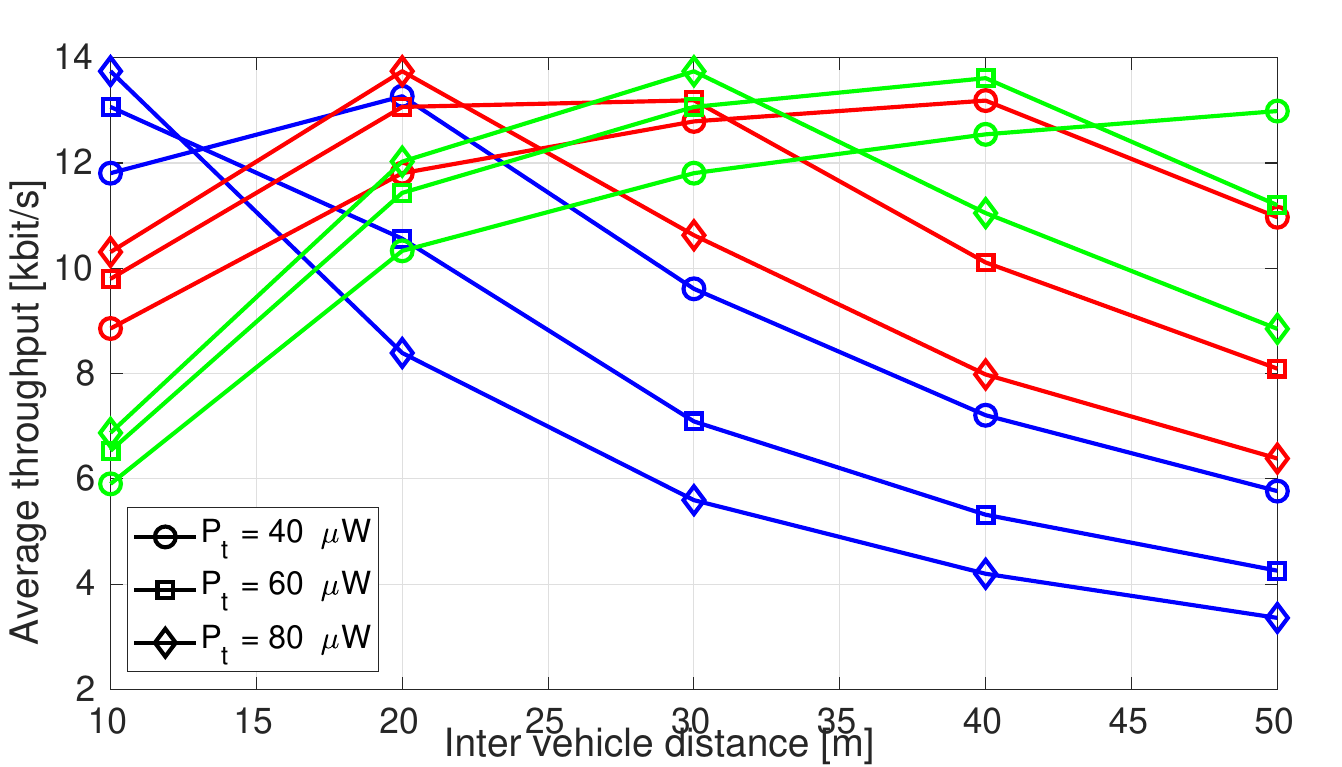}
 \caption{\small Average throughput for varying values of the inter vehicle distance $\dpla$. Blue lines are for $\ell=1$ m, red lines for $\ell = 2$ m and green lines for $\ell = 3$ m.}
 \label{fig:vardv_Pt_Fix}
 \vspace{-0.5cm}
 \end{center}
\end{figure}

In the high vehicular traffic scenario, $\dpla$ can be, to a certain extent, also tuned by the EHD, by choosing to harvest energy only from a subset of vehicles. Figure~\ref{fig:vardv_Pt_Fix} plots the throughput as a function of $\dpla$ in this scenario, with $S=2$ kbit.
First of all, we observe that for any given values of $\dpla$ and $\ell$, there is an optimal transmit power level, which is the one that fully balances the decoding probability at the AP and the energy outage probability at the EHD. Secondly, if the vehicular traffic is high enough ($\dpla=10$ m), the best option is to set $P_t$ to its maximum value and $\ell$ to a low value. As the traffic becomes less congested, larger values of $\ell$ are preferable, since energy sources are available less frequently, and increasing $\ell$ can prevent throughput losses.
However, when $\dpla$ further increases, lowering $P_t$ becomes necessary. Notice that almost the same throughput (between 13 and 14 kbit/s) is achievable over the entire span of $\dpla$ values, meaning that a proper selection of the transmission and harvesting parameters can grant stable performances under varying vehicular traffic conditions.

\begin{figure}
 \begin{center}
   \includegraphics[width=\figww]{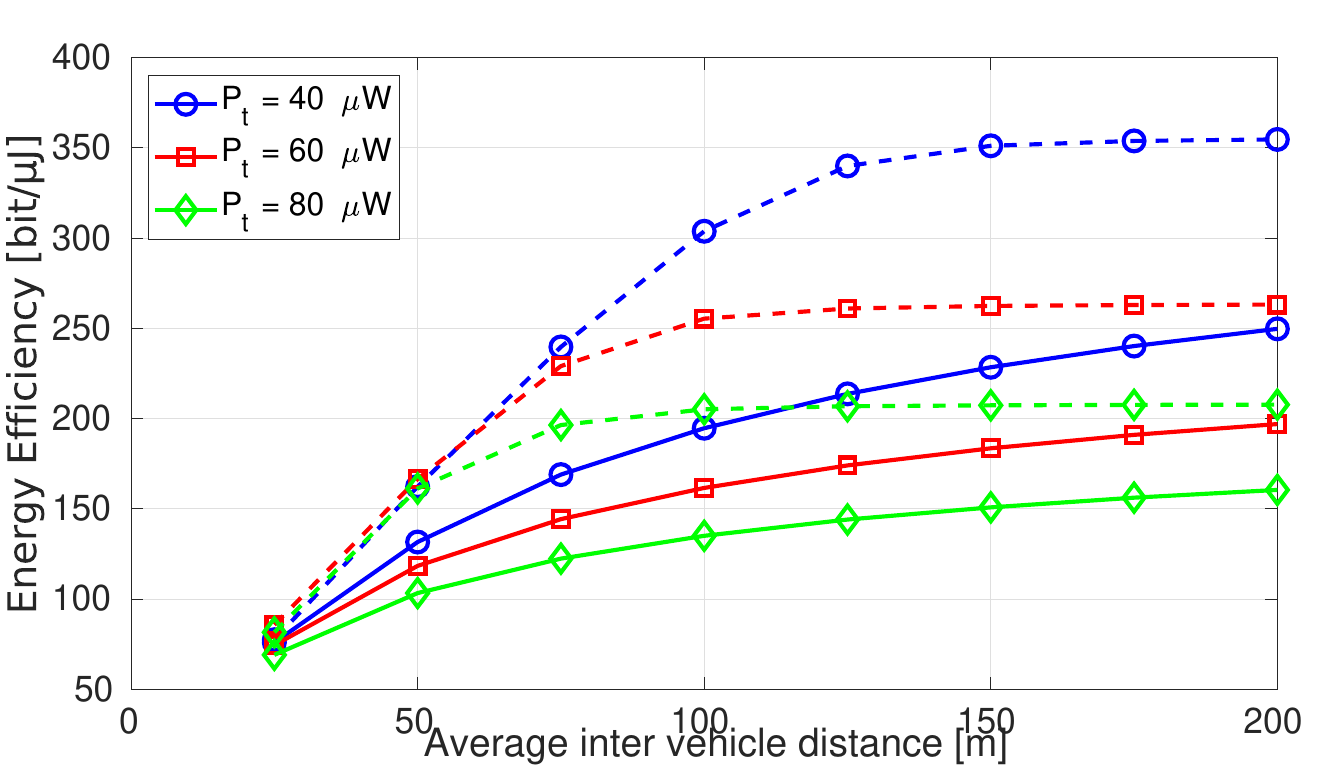}
 \caption{\small Energy efficiency in the platooning scenario (dashed lines) and in the sparse traffic scenario (continuous lines).}
 \label{fig:vardv_PlatGain}
 \vspace{-0.5cm}
 \end{center}
\end{figure}

We conclude this section by comparing the energy efficiency $\Upsilon$ in the two traffic scenarios, as a function of the traffic density.
Figure~\ref{fig:vardv_PlatGain} depicts, for different values of the transmit power, the maximum energy efficiency attained in the platooning scenario and in the sparse traffic scenario, each one computed for the value of $\ell$ that maximizes the expected throughput. The packet size $S$ is set to 2 kbit.
Energy efficiency initially grows with the average inter-vehicle distance $\mathbb{E}[d_v]$, since energy overflow events, which bring to energy waste, become less frequent. However, as $\mathbb{E}[d_v]$ further increases, it finally reaches an upper limit, which corresponds to the case when vehicles are so distant that the entire energy harvested from each one is completely used before a new cycle begins.

While this trend holds for both scenarios, we observe that platooning is able to grant a much faster growth, and a higher energy efficiency, with a top gain vs irregular traffic of more than 55\%.
Indeed, a regular traffic pattern enables a better energy utilization: knowing that a new incoming energy source lies at a predefined distance allows the system to make the best usage of the energy in the battery, with little concern for an energy outage. Conversely, stochastic vehicles arrivals force the system to overprovision energy in order to face longer intervals between successive battery recharges. We can call this effect the \emph{price of uncertainty}, which is relevant especially when the inter vehicle distance is not too large and the transmit power is high. 

\subsection{Black-out probability}
While throughput is a good metric to evaluate the average data rate received at the AP, it cannot fully capture the overall communication quality. Indeed, Age of Information (AoI) is another relevant quantity to be analyzed. It is especially important for some safety applications, which require timely information updates in order to trigger proper recovery actions.
In particular, the so called \emph{Black out} events~\cite{myVT}, that is, AoI exceeding a predefined value $Q_s$, can be potentially harmful, and should be avoided. While the proposed cycle-based harvesting strategy is not specifically designed to minimize the black out probability, its analysis can shed light on the best parameters setup to limit its value.
The analysis is particularly interesting in the high vehicular traffic scenario (in the sparse traffic one, most of the black outs are likely to occur when two subsequent vehicles are very far from each other, so there is simply no energy to perform transmissions, regardless of the adopted strategy), where a closed form expression can be found (see Appendix \ref{app:probo}). 

\begin{figure}
 \begin{center}
   \includegraphics[width=\figww]{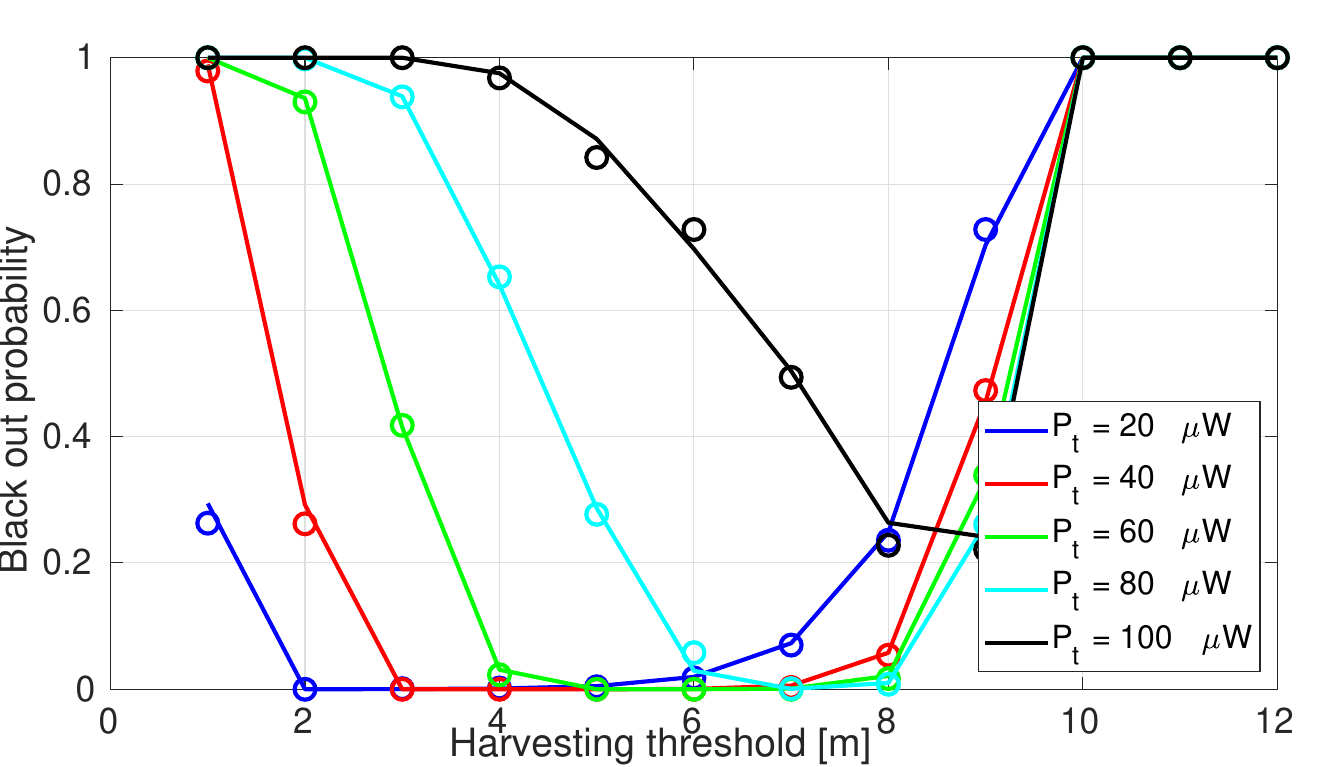}
 \caption{\small Black out probability for varying values of the transmit power $P_t$. Continuous lines are theoretical results, markers are for simulation results.}
 \label{fig:BO_varlim_Pt_Fix}
 \vspace{-0.5cm}
 \end{center}
\end{figure}
In Figure~\ref{fig:BO_varlim_Pt_Fix}, we set $Q_s$ to 2 s, and the inter vehicle distance to 50 m. The black out probability in a cycle is plotted against the harvest distance $\ell$. Also for this metric, the theoretical results fit the simulation curves very well. For all the transmit power levels, the curve shows a minimum, and then goes to 1 for $\ell=10$ m. This is not surprising, since at this value the duration of the HP is already equal to 2 s, and a black out occurs at every cycle.
However, a too low value of $\ell$ is also a bad choice: while the HP causes only a short transmission interruption, the harvested energy is not enough to sustain the data communications until the end of the cycle, and the resulting energy outage turns into a black out event.
Clearly, the optimal $\ell$ grows with the transmit power, but rising $P_t$ too much makes it impossible to lower the black out probability to almost 0.

\begin{figure}
 \begin{center}
   \includegraphics[width=\figww]{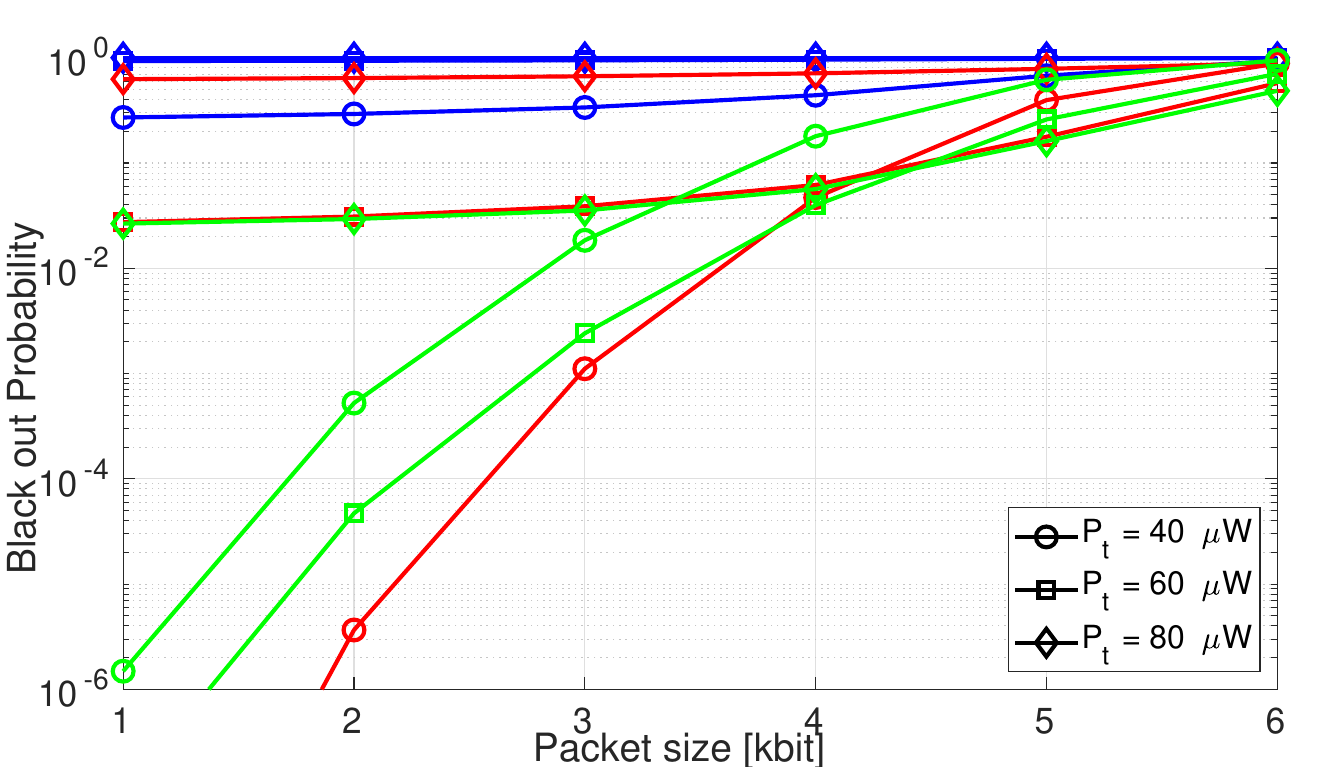}
 \caption{\small Black out probability for varying values of the transmit power $P_t$. Blue lines are for $\ell=2$ m, red lines are for $\ell=4$ m, and green lines are for $\ell=6$ m. The inter vehicle distance is $\dpla = 50$ m.}
 \label{fig:BO_varlen_Pt_Fix}
 \vspace{-0.5cm}
 \end{center}
\end{figure}
The black out probability as a function of the data rate is illustrated in Figure~\ref{fig:BO_varlen_Pt_Fix}.
As expected, it increases with the data packet size $S$, and $\ell=6$ m grants the best performance for $S\geq 4$ kbit.
By comparing this figure with Figure~\ref{fig:varlen_Pt_Fix}, an interesting tradeoff is unveiled.
Setting a large $\ell$ and a large $P_t$ grants the highest throughput for $S=4$ kbit, but only a relatively high black-out probability, around 0.05. Conversely, if a black-out probability of $10^{-3}$ is to be guaranteed, $P_t$ must be necessarily lowered. Setting $S$ to 3 kbit, with $\ell=4$ m, and $P_t=40$ $\mu$W achieves the desired black-out probability, at the cost of a 20\% throughput loss.
The best parameter tuning for the EHD, therefore, strongly depends on the requirements of the specific application implemented on the system.

\subsection{Impact of channel model}
We conclude the paper by analyzing the effect of different fading models over the two main performance metrics.
\begin{figure}
 \begin{center}
   \includegraphics[width=\figww]{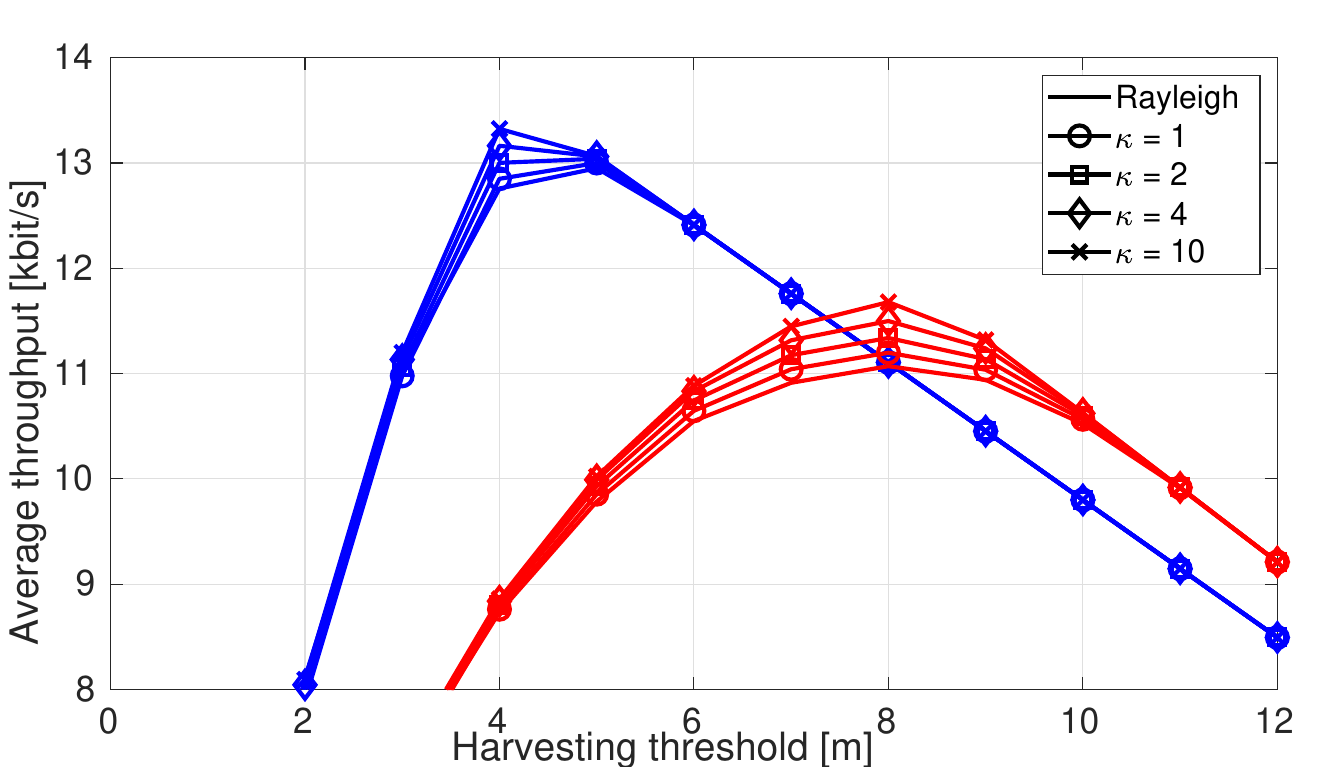}
 \caption{\small Average throughput for different channel models. Blue lines are for $P_t=60$ $\mu$W, red lines are for $P_t = 100$ $\mu$W.}
 \label{fig:varcha_Pt_Fix}
 \vspace{-0.5cm}
 \end{center}
\end{figure}
\begin{figure}
 \begin{center}
   \includegraphics[width=\figww]{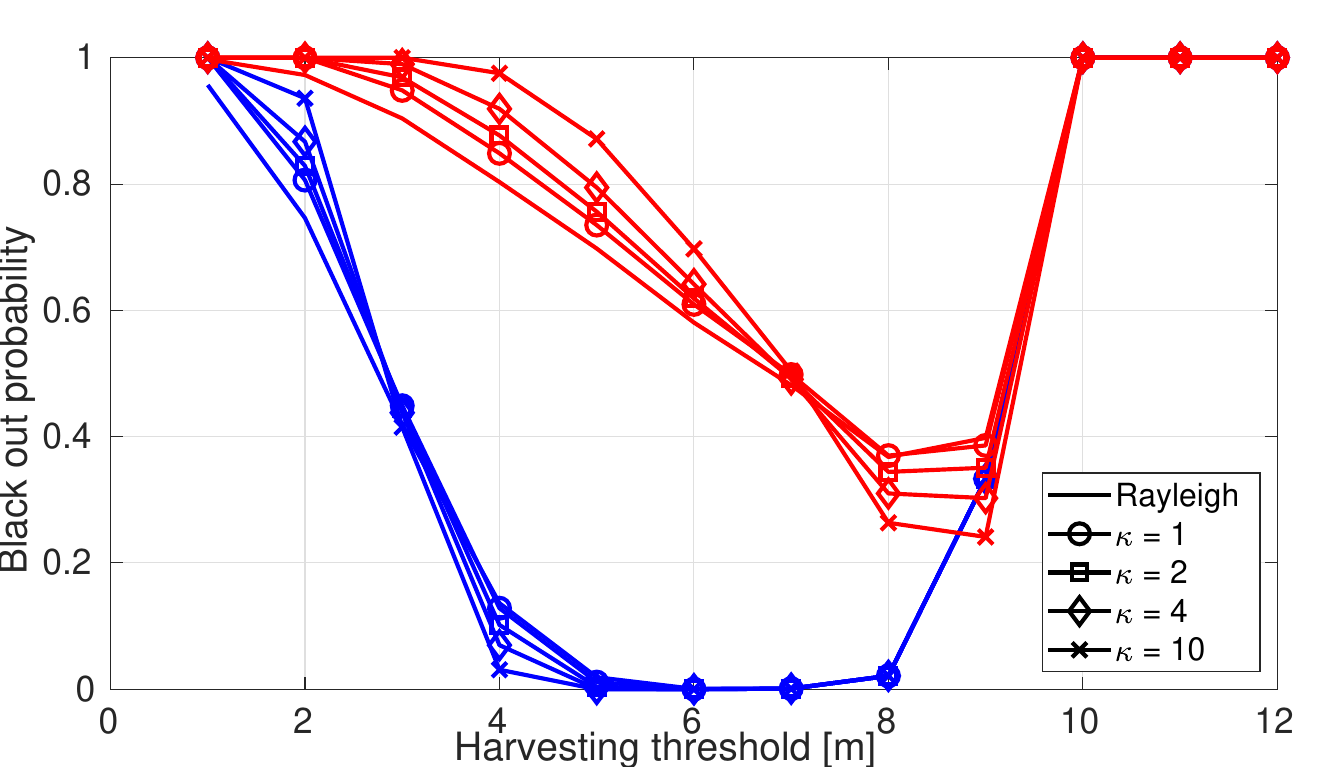}
 \caption{\small Black out probability for different channel models. Blue lines are for $P_t=60$ $\mu$W, red lines are for $P_t = 100$ $\mu$W.}
 \label{fig:BO_varcha_Pt_Fix}
 \vspace{-0.5cm}
 \end{center}
\end{figure}
In Figure~\ref{fig:varcha_Pt_Fix} we plot the average throughput as a function of the harvesting threshold, for two values of $P_t$, comparing the Rician fading model with different values of the Rice factor $\kappa$, and also the Rayleigh model (for which the throughput derivation is reported in Appendix~\ref{app:ray}). The black-out probability in the same scenario is instead depicted in Figure~\ref{fig:BO_varcha_Pt_Fix}. As to the throughput, we observe that Rician fading offers better performance than Rayleigh fading, although this is relevant only over a limited interval. This is because the average value of the harvested energy over a cycle with both models is the same, but the variance is different: Rayleigh fading (and in general, lower values of $\kappa$), gives in fact a larger variance. When the average harvested energy brings the battery level close to its capacity, however, this larger variance actually leads to an overall lower amount of scavenged energy. In fact, positive peaks of harvested energy, now capped by the finite battery capacity, do not compensate negative peaks, with a net effect of less scavenged energy.
Notice that this effect does not appear when the average amount of harvested energy in a cycle is quite low, and thus far from filling the battery (low values of $\ell$), nor when it is very high, and thus saturates the battery (high values of $\ell$). Since higher Rice factors reduce the variance of the fading, the effect becomes more evident for $\kappa=10$.

As to the black out probability, it is interesting to notice that Rayleigh fading can reduce it, with respect to Rician fading, up to a certain value of $\ell$, beyond which Rician fading gives better perfomance. This is again due to the lower variance of harvested energy per cycle given by Rician fading. Indeed, even if the average harvested energy per cycle is not enough to prevent black-out events (low $\ell$), Rayleigh fading is more likely to take values much greater than the average, thus making it possible to avoid some of the black-out events. Conversely, when the average harvested energy per cycle is high enough to ensure continuous data reception (high $\ell$), Rayleigh fading can still exhibit more pronounced negative peaks, thus causing black-out events. As before, since higher Rice factors reduce the fading variance, this effect is more evident when comparing Rayleigh fading and Rice fading with high $\kappa$.

\section{Conclusions}
\label{sec:conclu}
This paper proposes a cycle-based strategy for exploiting RF energy of vehicular communications by means of devices located alongside the road. A theoretical derivation of the average throughput is presented, and used to investigate the impact of several parameters, ranging from vehicular traffic type and intensity to transmit power, from data rate to battery capacity. Energy efficiency and black out probability are also derived.
Results show that the best performance is obtained with regular vehicular traffic patterns (e.g., platooning), and that transmit power and harvesting phase duration can be tuned in order to favor throughput maximization or black out probability minimization.
Furthermore, we observe that the presence of regular traffic patterns can substantially increase the achievable throughput and the energy efficiency by reducing the price of uncertainty. This is of particular relevance, since recent or novel technologies like V2X communications, autonomous vehicles or smart intersections, are expected to enhance coordination among vehicles, in order to reduce (electric) fuel consumption, thus leading to more predictable traffic patterns. A major implication of our study is that platooning is useful also for providing a much more effective source of RF energy to surrounding wireless networks based on EH, thus highlighting for the first time an additional potential benefit of platooning applications.

Three potential directions for future work are envisioned. Firstly, the performance in more complex scenarios, with multiple lanes or roads, may offer additional insights; secondly, the development of a non agnostic, more adaptive strategy, able to leverage the knowledge of the battery status, appears to be promising; thirdly, an experimental validation, even on a simple case study, may assess the feasibility of the illustrated framework, and pave the way to more refined approaches.

\begin{appendices}
\section{Throughput derivation with Rayleigh fading}
\label{app:ray}
With Rayleigh fading, $h_v(n)$ is modeled as a complex Gaussian random variable with zero mean and unit variance. Equation (\ref{segdiv}) still holds, but now the $X_i$'s have an exponential distribution with unitary mean. Therefore, the random variable $Y_i = X_i/\lambda_i$ is still exponential, with mean $1/\lambda_i$, and $E_h$ is the sum of exponential random variables with different parameters.
Let us divide the sum in (\ref{segdiv}) into two parts, namely
\begin{equation}
 E_h^{(1)} = \sum_{i=0}^{L/2-1}Y_i,\quad\quad E_h^{(2)} = \sum_{i=L/2}^{L-1}Y_i.
\end{equation}
The former sum corresponds to the approaching of the vehicle $V_q$ that provides energy, from point $(\xehd-\ell,0)$ to point $(\xehd,0)$, the latter instead to the interval when $V_q$ moves away, from point $(\xehd,0)$ to point $(\xehd+\ell,0)$, as per Figure~\ref{fig:gramodel}.

Let us consider the approaching phase. Since all the $\lambda_i$'s are different, and leveraging the independence of the $X_i$'s, the probability density function of $E_h^{(1)}$ is modeled as an hypoexponential distribution, that is
\begin{equation}
 f_{E_h}^{(1)}(x) = \left(\prod_{i=0}^{L/2-1}\lambda_i\right)\sum_{j=1}^K\frac{e^{-\lambda_jx}}{\prod_{k\neq j}(\lambda_k-\lambda_j)}.
 \label{pdfE}
\end{equation}

For symmetry reasons, the energy $E_h^{(2)}$ harvested when the vehicle is moving away has exactly the same distribution. Therefore, the overall energy $E_h$ harvested in the entire HP is $E_h^{(1)}+E_h^{(2)}$, whose cumulative distribution function is given by
\begin{equation}
 F_E(x) = \int_0^{x}f_{E_h}^{(1)}(y)F_{E_h}^{(1)}(x-y)\de y,
\end{equation}
where the CDF $F_{E_h}^{(1)}(x)$ of $E_h^{(1)}$ is easily obtained from (\ref{pdfE}) through integration. After mathematical manipulations,
we get
\begin{eqnarray}
 F_E(x) & = & 1 - A\sum_{i=0}^{L/2-1}\left(\sum_{j\neq i}\frac{e^{-\lambda_jx}/\lambda_j -e^{-\lambda_ix}/\lambda_i}{(\lambda_i-\lambda_j)\Lambda_i\Lambda_j}+\right. \nonumber \\
 & & \left. + \frac{e^{-\lambda_ix}(1+\lambda_ix)}{\lambda_i^2\Lambda_i^2}\right.\Bigg)
 \label{ecdf}
\end{eqnarray}
where $A=\prod_{i=0}^{L/2-1}\lambda_i^2$ and $\Lambda_i=\prod_{k\neq i}(\lambda_k-\lambda_i)$.
We can put the obtained CDF into (\ref{poithro}) and into (\ref{throplat}) to obtain the average conditioned number of transmissions per cycle in the case of sparse and heavy vehicular traffic, respectively.

\section{Derivation of the Black-out Probability}
\label{App}
\label{app:probo}

\begin{figure}
 \centering
 \begin{tikzpicture}[>=stealth, scale=.6]
  \def\lato{0.5}; 
  \foreach \vc in {0,1,...,5}
  {
    \filldraw[draw=black, fill=cyan] (\vc*\lato,0) rectangle +(\lato,\lato);
  }
  \foreach \vc in {16,17,...,21}
  {
    \filldraw[draw=black, fill=cyan] (\vc*\lato,0) rectangle +(\lato,\lato);
  }
  \foreach \vc in {6,7,...,13}
  {
    \filldraw[draw=black, fill=green] (\vc*\lato,0) rectangle +(\lato,\lato);
  }
  \foreach \vc in {22,23,...,25}
  {
    \filldraw[draw=black, fill=green] (\vc*\lato,0) rectangle +(\lato,\lato);
  }
  \foreach \vc in {14,15,...,15}
  {
    \filldraw[draw=black, fill=yellow] (\vc*\lato,0) rectangle +(\lato,\lato);
  }
  \foreach \vc in {11,12,...,13}
  {
    \node at (\vc*\lato+\lato/2,\lato/2) {?};
  }
  \foreach \vc in {22,23,...,24}
  {
    \node at (\vc*\lato+\lato/2,\lato/2) {?};
  }
  
  \draw[->,thick] (-0.5,0) -- (26*\lato+0.5,0) node[right,yshift=-.3cm,xshift=-.3cm] {$t$};
  
  \draw[thick] (0,0) -- (0,2);
  \draw[thick] (16*\lato,0) -- (16*\lato,2);
  \draw[thick] (6*\lato,0) -- (6*\lato,\lato+1);
  
  \draw[<->, thick] (0,\lato+1.4) -- (16*\lato,\lato+1.4) node[midway,yshift=0.3cm] {Cycle ($N$ slots)};
  \draw[<->, thick] (0,\lato+0.5) -- (6*\lato,\lato+0.5) node[midway,yshift=0.3cm] {\footnotesize HP ($N_H$ slots)};
  \draw[<->, thick] (6*\lato,\lato+0.5) -- (16*\lato,\lato+0.5) node[midway,yshift=0.3cm] {\footnotesize TP ($N_T$ slots)};
  \draw[decorate,decoration={brace,amplitude=5pt},xshift=0pt,yshift=0pt] (14*\lato,-0.1) -- +(-8*\lato,0) node[black,midway,yshift=-0.4cm]{$\Ntx$};
  \draw[decorate,decoration={brace,amplitude=5pt},xshift=0pt,yshift=0pt] (16*\lato,-0.1) -- +(-2*\lato,0) node[black,midway,yshift=-0.4cm]{$\Nno$};
 \end{tikzpicture}
 \caption{Illustration of the occurrence of a potential blackout of $x=12T$ seconds, corresponding to 11 missed transmissions. In this case, the battery charge at the end of the first $N_H$ slots allows $\Ntx=8$ transmissions, while in the last $\Nno=2$ slots of the TP no transmission can be performed.
 Therefore, an interruption of $N_H+\Nno=8$ slots occurs. The probability that a black out of 11 consecutive slots happens is the probability that there are at least 3 consecutive failures in the set of 6 slots marked with a ``?'' in the figure.}
 \label{fig:diagpo}
 \vspace{-.3cm}
\end{figure}
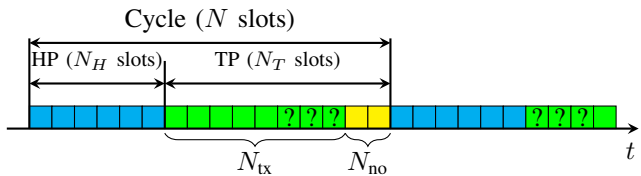
We compute the expression for the black out probability $\Pbo(x)$, defined as the probability that the AoI equals or exceeds the time duration of $x$ time slots in a cycle. This occurs when no successful transmission is performed over at least $x-1$ consecutive time slots.

In the high traffic scenario, with fixed inter vehicle distance $d_v=\dpla$, the number of time slots per cycle is given by $N = \dpla/(v_0T)$. Among these, $N_H = 2\ell/(v_0T)$ slots are reserved to the harvest phase, while the remaining $N_T = N-N_H$ are reserved for transmissions.
It follows that, in a given cycle, there are at least $N_H$ consecutive slots where no transmission is performed. Therefore, if $x\leq N_H$, we have $\Pbo(x) = 1$.

Let us now move to the more interesting case when $x > N_H$.
In a general cycle, the $N_T$ slots of the transmit phase can be further divided into two groups. In the former group of $\Ntx$ slots, the EHD has enough energy to transmit, whereas in the subsequent $\Nno = N_T-\Ntx$ slots, no transmission is performed due to energy outage (see Figure~\ref{fig:diagpo}). Clearly, it can be $\Ntx=N_T$ (and hence $\Nno=0$) if the battery has enough charge.

We state the following assumptions:
\begin{itemize}
 \item $\Ntx>0$, meaning that after the harvest phase, the battery has enough energy to perform at least one transmission. This is almost always true, since the energy $\Etx$ required to send a single data packet is much lower than the battery capacity, and $F_E(\Etx)\simeq 0$.
 \item the longest interval with no decoded packets in a cycle contains the $N_H$ slots of the HP and the last $\Nno$ slots of the TP. In fact, in these slots no transmissions occur, and we only need $x-(N_H+\Nno)$ consecutive failed transmissions to have a black out. Since $N_H+\Nno>0$, this event has a much higher probability than having $x$ consecutive failed transmissions among the $\Ntx$ performed ones.
\end{itemize}

Under these assumptions, we can condition the black out probability over the distribution $\hat p_B$ of the battery status \emph{after} the harvesting phase:
\begin{equation}
 \Pbo(x) = \sum_{i=0}^{N_s}\Pbo(x|i)\hat p_B(i\Etx),
 \label{probo}
\end{equation}
We can easily find $\hat p_B(i\Etx)$ as
\begin{equation}
 \hat p_B(i\Etx) = \sum_{j=0}^ip_B(j\Etx)p_E(i-j),
\end{equation}
where $p_B$ is the pmf of the battery status before the HP, obtained in Section~\ref{sec:batstat}, and $p_E$ is the pmf of the energy quanta harvested during the HP, defined in (\ref{quapdf}).

The quantized battery status corresponds to the number $\Ntx$ of transmissions that can be supported in the cycle. In other words, $\Pbo(x|i)$ is the black out probability given that $\Ntx=i$.
We can distinguish two cases.
If $i \leq N-x+1$, then there is not enough energy in the battery to avoid an interruption of $x-1$ slots (including the $N_H$ ones), and the black out occurs with probability $\Pbo(x|i)=1$.
If instead $i \geq N-x$, then there is enough energy to avoid the black out, and its occurrence depends only on decoding failures.
The number of silent slots where transmissions do not take place is $N_H+\Nno$, which is equal to $N-i$. In order to have a ``hole'' of $x-1$ slots, we need at least $w = x-1-(N-i)$ consecutive failures in the slots adjacent to the silent slots, as sketched in Figure~\ref{fig:diagpo}.
The probability that this happens can be found by considering only the 
$w$ slots before the silent slots, and the $w$ slots after them. If a sequence of at least $w$ consecutive failures can be found within these $2w$ slots, the black out occurs\footnote{We are implicitly assuming that there is no energy outage in the first $w$ slots after the HP. While this is not always true, its probability is high, as long as $w$ is limited.}.
The probability $\varphi(w,q)$ of having at least $w$ consecutive successes among a sequence of $2w$ i.i.d. attempts, with success probability $q$, is obtained by summing the probabilities of having exactly $j$ consecutive successes, with $w\leq j\leq2w$; each of them can be conditioned on the total number $k\geq j$ of successes, which follows a binomial distribution, thus yielding
\begin{equation}
 \varphi(w,q) = \sum_{j=w}^{2w}\sum_{k=j}^{2w}Q(2w,j|k)\binom{2w}{k}q^k(1-q)^{2w-k},
\end{equation}
The term $Q(L,j|k)$ is the probability of having $j$ consecutive successes over $L$ attempts, given that there are $k\geq j$ successes. For the case of interest $j\geq L/2$, its value is
equal to 0 when $j<k=L$, to 1 when $j=k=L$, and to $2/L$ if $k=L-1$; otherwise, its expression is computed as
\begin{equation}
 Q(L, j| k) = \frac{2\binom{L-j-1}{k-j} + (L-j-1)\binom{L-j-2}{k-j}}{\binom{L}{k}}.
\end{equation}

The final expression for the conditioned black out probability therefore reads as
\begin{equation}
 \Pbo(x|i) = 
 \begin{cases}
  1 & \text{if } i\leq N-x+1, \\
  \varphi(i-N+x-1,1-\phi_s) & \text{otherwise},
 \end{cases}
\end{equation}
where $\phi_s$ is the packet decoding probability in (\ref{decprob}). This expression can be put into (\ref{probo}) to get the overall black out probability in a cycle.

\end{appendices}

\bibliographystyle{IEEEtran}
\bibliography{IEEEabrv,biblio}

\end{document}